\documentclass[aps,preprint,showpacs,preprintnumbers]{revtex4-1}
\usepackage{epsfig}
\usepackage{latexsym,amsmath,amssymb,amstext}
\usepackage{float}
\usepackage{graphicx,xcolor}
\usepackage[left=0.85in,right=0.85in,top=0.9in,bottom=0.9in]{geometry}

\newcommand{\be}{\begin{equation}}
\newcommand{\ee}{\end{equation}}
\newcommand{\bea}{\begin{eqnarray}}
\newcommand{\eea}{\end{eqnarray}}

\newcommand{\feyn}[1]{#1\!\!\!\!\slash\  }

\newcommand\bef{\begin{figure}}
\newcommand\eef[1]{\label{fg:#1}\end{figure}}
\newcommand\beq{\begin{equation}}
\newcommand\eeq[1]{\label{#1}\end{equation}}
\newcommand\beqa{\begin{eqnarray}}
\newcommand\eeqa[1]{\label{#1}\end{eqnarray}}
\newcommand\bet{\begin{table}}
\newcommand\eet[1]{\label{tb:#1}\end{table}}

\newcommand\fgn[1]{Figure \ref{fg:#1}}
\newcommand\eqn[1]{Eq.\ (\ref{#1})}
\newcommand\scn[1]{Section \ref{sec:#1}}
\newcommand\apx[1]{Appendix \ref{sec:#1}}
\newcommand\tbn[1]{Table \ref{tb:#1}}

\newcommand\ie{{\sl i.e.\/}}

\begin{document}
\title{No evidence for bilinear condensate in parity-invariant three-dimensional QED with massless fermions}
\author{Nikhil\ \surname{Karthik}}
\email{nkarthik@fiu.edu}
\affiliation{Department of Physics, Florida International University, Miami, FL 33199.}
\author{Rajamani\ \surname{Narayanan}}
\email{rajamani.narayanan@fiu.edu}
\affiliation{Department of Physics, Florida International University, Miami, FL 33199.}

\begin{abstract}
We present our numerical study of three-dimensional QED with 2, 4, 6 and
8 flavors of massless two-component fermions using a parity-preserving
 lattice regularization with Wilson fermions. We study the behavior
of low-lying eigenvalues of the massless improved Wilson-Dirac operator as a
function of three-dimensional physical volume, after taking the continuum
limit at fixed physical volumes. We find the following evidences against
the presence of bilinear condensate:  the eigenvalues do not scale as
the inverse of the three-dimensional physical volume, and the number
variance associated with these eigenvalues do not exhibit ergodic
behavior. The inverse participation ratio (IPR) of the associated
eigenvectors exhibits a multi-fractal volume scaling. The  relation
satisfied by number variance and IPR suggests critical behavior.
\end{abstract}

\date{\today}
\maketitle

\section{Introduction}

Two component massless fermions coupled to a three-dimensional Euclidean
abelian gauge field has been a topic of study in the past three decades
 for several field-theoretic reasons, and it is also of interest  to
condensed matter physics~\cite{Miransky:2015ava}. The massless Dirac
operator is
\be
\feyn{ C}({\bf A}) =  \sum_{k=1}^3 \sigma_k \left[ \partial_k + i A_k({\bf x})\right],
\ee
where $\sigma_k$ are the Pauli matrices and $A_k({\bf x})$ is a background
abelian field. Under parity,
\be
{\bf x} \to  -{\bf x}; \qquad
{\bf A}({\bf x}) \to  -{\bf A}(-{\bf x}); \qquad
\feyn{C}({\bf A}) \to  \feyn{C}^\dagger({\bf A})  = -\feyn{C}({\bf A}).
\ee
This theory has a parity anomaly~\cite{Deser:1981wh,Deser:1982vy} since
the fermion determinant is not real~\cite{Niemi:1983rq,Redlich:1983dv}
and its parity-violating phase is regulator
dependent~\cite{So:1984nf,So:1985wv,Coste:1989wf}. Furthermore, the form
of the parity-violating term at finite temperature and non-trivial gauge
field backgrounds can be quite different from the perturbative infinite
volume result~\cite{Deser:1997gp,Karthik:2015sza}.

This parity anomaly can be cancelled by suitable regularization when even
number of flavors of two-component  fermions are present. In this paper, we 
are only interested in such a system. Assuming we
have a regulated version of $\feyn{C}({\bf A})$, we can write down a
parity invariant fermionic action for a $2N_f$ flavor theory as
\be
S_f = \int d^3 x \sum_{i=1}^{N_f} 
\left\{ \bar\chi_i({\bf x}) \feyn{C}({\bf A}) \chi_i({\bf x})
+\bar\phi_i({\bf x}) \feyn{C}^\dagger({\bf A}) \phi_i({\bf x})
\right\},
\ee
with the fermions transforming under parity as
\be
\bar\chi_i({\bf x}) \to  \bar\phi_i({\bf x}); \qquad
\chi_i({\bf x}) \to  \phi_i({\bf x}).
\ee
It is useful to identify the $2N_f$ flavors of 2-component fermions
as $N_f$ flavors of  4-component fermions using the following
notation~\cite{Pisarski:1984dj},
\be
\psi_i({\bf x}) = \begin{pmatrix} \phi_i({\bf x}) \cr \chi_i({\bf x}) \end{pmatrix};\qquad
\bar\psi_i({\bf x}) = \begin{pmatrix} \bar\chi_i({\bf x}) & \bar\phi_i({\bf x}) \end{pmatrix};\qquad
\feyn{D}({\bf A}) = \begin{pmatrix} 0 & \feyn{C}({\bf A}) \cr \feyn{C}^\dagger({\bf A}) & 0 \end{pmatrix},
\label{diracop}
\ee
with the associated action,
\be
S_f = \int d^3 x \sum_{i=1}^{N_f} 
\bar\psi_i({\bf x}) \feyn{D}({\bf A}) \psi_i({\bf x}).
\label{action}
\ee
Note that the usual anti-hermitian Dirac operator is
\be
\gamma_5 \feyn{D}({\bf A}) = \sum_{k=1}^3 \gamma_k \left[ \partial_k + i A_k({\bf x})\right].
\ee
Under parity,
\be
\psi_i({\bf x}) \to  P \psi_i({\bf x}); \qquad
\bar\psi_i({\bf x}) \to  \bar\psi_i({\bf x}) P;\qquad
P=\begin{pmatrix} 0 & 1 \cr 1 & 0 \end{pmatrix};
\qquad
\feyn{D}({\bf A}) \to P \feyn{D}({\bf A}) P = -\feyn{D}({\bf A}).
\ee

The action in \eqn{action} has a U($2N_f$) symmetry. The two fermion
bilinears which break the symmetry to U$(N_f)\times$U$(N_f)$, but 
invariant under parity, are
\be
S_m = \int d^3 x \left[ m_p \bar\psi_i({\bf x}) P \psi_i({\bf x})
+ im \bar\psi_i({\bf x}) \psi_i({\bf x}) \right].\label{mmp}
\ee
The fermion determinant in a fixed gauge field background becomes
\be
Z(m,m_p) = \det \left (\feyn{C}({\bf A})\feyn{C}^\dagger({\bf A}) + m^2 + m_p^2\right),
\ee
which makes different choices of $m$ and $m_p$ equivalent as long as
$m^2+m_p^2$ remains the same. These different ways to  introduce mass
will be used advantageously in our lattice formulation. If the U$(2N_f)$
symmetry gets spontaneously broken to U$(N_f)$ $\times$ U$(N_f)$ in the massless theory,  then
$\bar\psi_i\psi_i$ and $\psi_i P\psi_i$ will pick a vacuum expectation
value, which we  refer to as the bilinear condensates. With the right
ordering of limits (the three-dimensional volume $\ell^3$ is taken to infinity
before the fermion mass is taken to zero),  the bilinear condensates are
\be
\lim_{m\to 0} \lim_{\ell\to\infty}\frac{1}{\ell^3}\frac{\partial \log \left\langle Z(m,0)\right\rangle}{\partial m} \ne 0\qquad {\rm and}\qquad
\lim_{m_p\to 0} \lim_{\ell\to\infty}\frac{1}{\ell^3}\frac{\partial \log\left\langle Z(0,m_p)\right\rangle}{\partial m_p} \ne 0.
\ee
Analytic arguments in favor of a non-zero bilinear condensate were
first provided by Pisarski~\cite{Pisarski:1984dj} in the limit of
large number of flavors. The associated gap equation was analyzed
in~\cite{ Appelquist:1985vf, Appelquist:1986fd, Appelquist:1986qw,
Appelquist:1988sr} which supports a non-zero bilinear condensate if $N_f <
4$. It is worthwhile noting that the massless fermion propagator was used
in the fermion bubbles summed up to obtain the gauge boson propagator.
In addition, the wave function renormalization of the fermion was set to
unity in the limit of large $N_f$. Since the computation was performed
in the large-$N_f$ limit and an upper bound on $N_f$ was obtained for a
non-zero bilinear condensate, a different approach is needed to verify
this result.  Furthermore, estimating the free energy by simply counting
the degrees of freedom in the UV and IR assuming a non-zero bilinear
condensate suggests a bound of $N_f < 2$~\cite{Appelquist:1999hr}. 
In another study~\cite{Giombi:2015haa}, the stability of the conformal 
fixed point of QED in $4-\epsilon$ dimensions on a sphere, 
extrapolated to three dimensions, suggests a critical $N_f < 4$.
A schematic phase diagram of QED$_3$ as a function of the number of flavors
is shown in \fgn{phasediag}, based on the above plausibility arguments. 
The question we try to answer, is the existence of the region with condensate and 
broken U$(2N_f)$ symmetry.
\bef
\begin{center}
\includegraphics[scale=0.7]{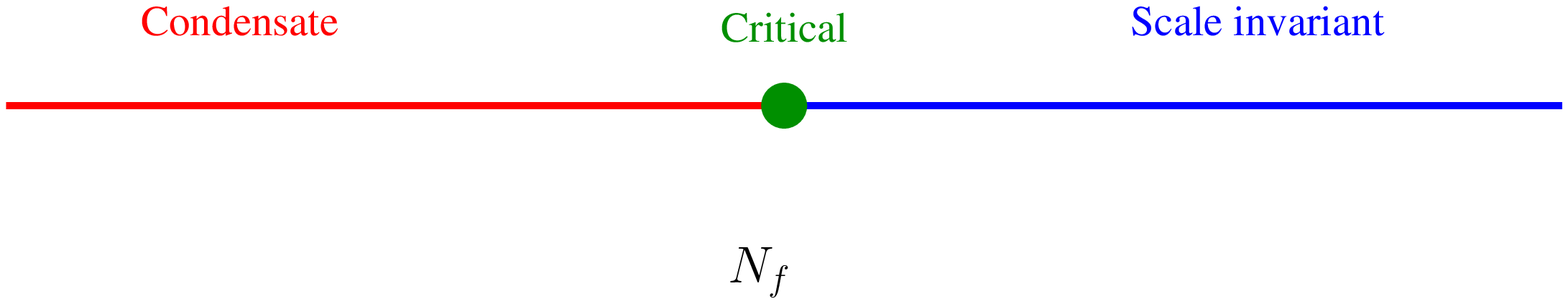}
\end{center}
\caption{The conjectured phase diagram of three-dimensional QED as
a function of the number of flavors $N_f$ of  massless 4-component
fermions. The blue region (to the right of critical point) has a
U$(2N_f)$ flavor symmetry and scale invariance. The symmetry is 
broken to U$(N_f)\times$U$(N_f)$ in the red region (to the left of
the critical point).  This paper deals with whether this region with
bilinear condensate (and  broken scale invariance) exists.
}
\eef{phasediag}

Extensive numerical studies have been carefully performed using staggered
fermions to investigate the possibility of a non-vanishing
 bilinear condensate. A single copy of staggered fermion results in a
 $N_f=2$ theory in the continuum. This particular example was studied
 in~\cite{Hands:2002dv} and an upper bound on the bilinear condensate
 was estimated.  By simulating a lattice model that uses the square
root of the staggered Dirac operator, the $N_f=1$ theory was carefully
 studied in~\cite{Hands:2004bh} and concluded that there is evidence
for a non-zero bilinear condensate in this theory.  In these studies,
 the theory with a fermion mass was simulated and the bilinear  was
measured as a function of mass to see if an extrapolation to zero mass
yielded a non-zero condensate in  the infinite volume limit. Although
several different physical volumes and lattice spacings were considered
 the analysis did not separate the two effects. In the $N_f=1$ case, 
increasing the physical volume at a fixed lattice spacing shows a trend
in the bilinear as a function of mass, that is in favor of a non-zero
condensate at zero mass. But, a comparison of two different physical
volumes (the largest two in the simulation) at two different lattice
spacings favors a vanishing condensate at zero mass. The $N_f=4$ theory
was also studied in~\cite{Hands:2004bh} with the aim of showing that
the condensate vanishes in this theory. An equation of state analysis of
the bilinear condensate as a function of fermion mass and lattice gauge
coupling does not convincingly provide evidence for a non-vanishing
condensate at $N_f=1$ nor for a vanishing condensate at $N_f=4$.

The aim of this paper is to revisit the problem numerically using Wilson
 fermions. The advantage of using Wilson fermions are two-fold. On the
one hand, we can simulate any value of $N_f$ without having to deal with
fractional powers of the lattice Dirac operator. On the other, there is a
place for both the bilinears in \eqn{mmp} --- we use $m_p$ on the lattice
to realize massless fermions and use $m$ to find evidence for a non-zero
condensate.   Wilson fermions were used earlier in~\cite{Raviv:2014xna}
to study the beta function of QED$_3$ with $N_f=2$;  here we use the two
different masses to study $N_f\ge 1$ and  explore fermionic observables.
In contrast to the studies in~\cite{Hands:2002dv,Hands:2004bh} where
one simulates a theory with a non-zero $m$, we simulate the theory with
$m=0$ and study the behavior of the low-lying eigenvalues as a function
of the physical volume. The low-lying eigenvalues were previously studied
 in the quenched approximation in \cite{Hands:1989mv}. Differing in the
 method,  our study here explicitly extracts the scaling behavior of
 the low-lying eigenvalues with respect to the volume. In particular,
 we expect the lowest eigenvalues of $\feyn{D}({\bf A})$, $\lambda_i$,
in a box of volume $\ell^3$, to scale such that the expectation value of
$\lambda_i \ell^3 \Sigma$ has a finite non-zero limit as $\ell\to\infty$
 with $\Sigma$ being the value of non-zero condensate.

The organization of the paper is as follows: we will present the 
parity-invariant formalism of Wilson fermions on the lattice in \scn{wil}.
We will present our results for $N_f=1$ in \scn{resn1} where we will
also make contact with certain ideas in random matrix theory in order
to understand the behavior of the low-lying modes of the improved Wilson-Dirac
operator. We will present the results for $N_f=2,3$ and 4 in \scn{resng1},
and compare them to the $N_f=1$ case.  This will be followed by our
conclusions.

\section{Parity invariant Wilson fermions}\label{sec:wil}

We used an isotropic $L^3$ lattice with periodic boundary conditions
in all three directions. Following~\cite{Hands:2002dv,Hands:2004bh},
we used the non-compact gauge action given by
\be
S_g=\frac{L}{\ell}\sum_n\sum_{j< k}^3 \left[\theta_j(n)+\theta_k(n+\hat j)-\theta_j(n+\hat k)-\theta_k(n)\right]^2,
\ee
where $\ell$ is the dimensionless linear extent of the periodic
box measured in units of the coupling constant. The  $\theta$'s are
related to the gauge-fields as $\theta_k=\frac{\ell}{L}A_k$. We used the
 Sheikhoslami-Wohlert-Wilson-Dirac operator~\cite{Sheikholeslami:1985ij}
 which was improved further by using one-level HYP smeared fields
 $\theta^s_k$~\cite{Hasenfratz:2001hp,Hasenfratz:2007rf} in the naive
 and Wilson terms. Let $C_W$ denote the two component
Sheikhoslami-Wohlert-Wilson-Dirac operator including the mass term
 $M_P$. We have given the details of smearing and $C_W$ in \apx{swwd}. In
 four component notation, our lattice realization of  the continuum 
 parity-invariant operator with mass terms (refer \eqn{action} and \eqn{mmp}) is
\be
 D_W = \begin{pmatrix} iM & C_W \cr C_W^\dagger & iM \end{pmatrix}.
\ee
The eigenvalues of $D_W$ come in complex conjugate pairs:
\bea
D_W \phi^\pm_j = \left(iM \pm \Lambda_j(M_P)\right) \phi^\pm_j;&\qquad&
\phi^\pm_j = \frac{1}{\sqrt{2}} 
\begin{pmatrix}
u_j \cr \pm \frac{1}{\Lambda_j(M_P)} C_W^\dagger u_j
\end{pmatrix},
\eea
where $\Lambda_j$ are the eigenvalues of the massless Dirac operator in lattice units.
In terms of the two-component Dirac operator,
\bea
C_W C_W^\dagger u_j = \Lambda_j^2(M_P) u_j;&\qquad& \Lambda_j(M_P) > 0;\qquad u_j^\dagger u_j =1.
\eea
As $L\to \infty$, the massless limit is obtained by setting $M=M_P=0$.
Due to the additive renormalization of $M_P$, one needs to tune $M_P$ as 
function of $L$
in order to remain massless after setting $M=0$. 

The dimensionless bilinear condensate is given by
\be
\Sigma = \frac{1}{\ell^2}\left\langle \sum_j \frac{2 m \ell}{\left(\lambda_j\ell\right)^2 + \left(m\ell\right)^2}\right\rangle,
\ee
where the quantities in the continuum are
\be
 \lambda_j \ell = \lim_{L\to\infty} \Lambda_j\big{(}M_P\big{)} L \qquad\text{and}\qquad m\ell=\lim_{L\to\infty} M L.
\ee
Then, for $\Sigma$ to be non-zero in the massless limit, we require the low-lying eigenvalues
to obey~\cite{Banks:1979yr,Verbaarschot:1994ip}
\be
 \lambda_j \ell = \frac{\Sigma z_j}{\ell^2},
\ee
with $z_j$ being certain universal numbers obtained
 from an appropriate random matrix theory that properly accounts for
the effective low energy Lagrangian with only the zero momentum mode
taken into account. This is an important requirement that we use to 
check if the bilinear condensate is present.

We explicitly set $M=0$ in our simulations. This enabled us to use the
standard Hybrid Monte Carlo~\cite{Duane:1987de} algorithm with $N_f$
copies of pseudofermions to simulate a $2N_f$ flavor theory. We
used $M_P$ to tune the theory to massless fermions.  The tuning was
achieved by  finding the $M_P$ that minimizes the lowest eigenvalue,
$\Lambda_1(M_P)$, over a small ensemble of thermalized configurations.
We computed these low-lying eigenvalues, including the associated
eigenvectors, using the Ritz algorithm~\cite{Kalkreuter:1995mm}. Our
simulation parameters are given in \apx{simpar}.

\section{Results for $N_f=1$}\label{sec:resn1}

\subsection{Effective potential using Wilson loops}

\bef
\begin{center}
\includegraphics[scale=0.9]{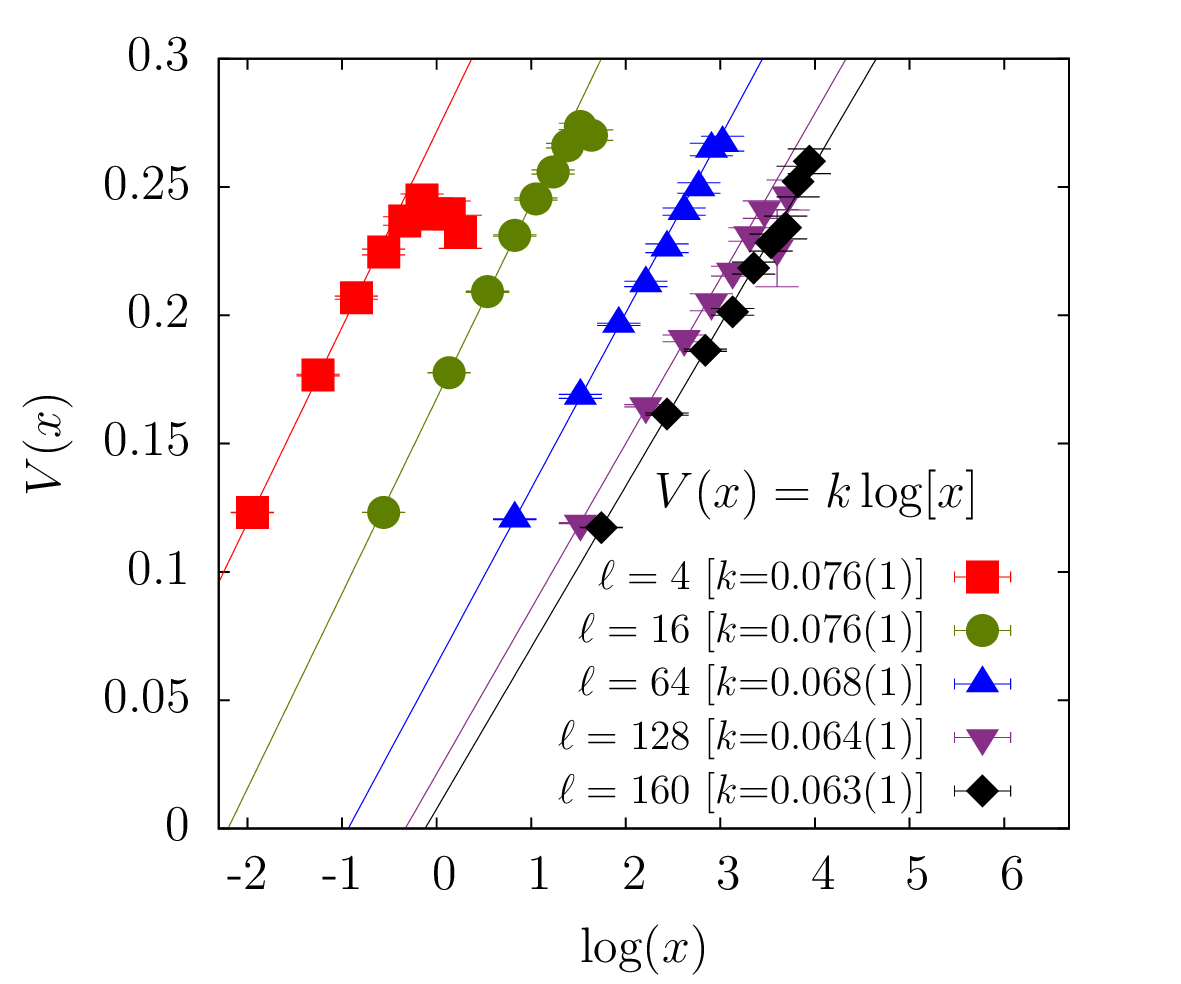}
\includegraphics[scale=0.9]{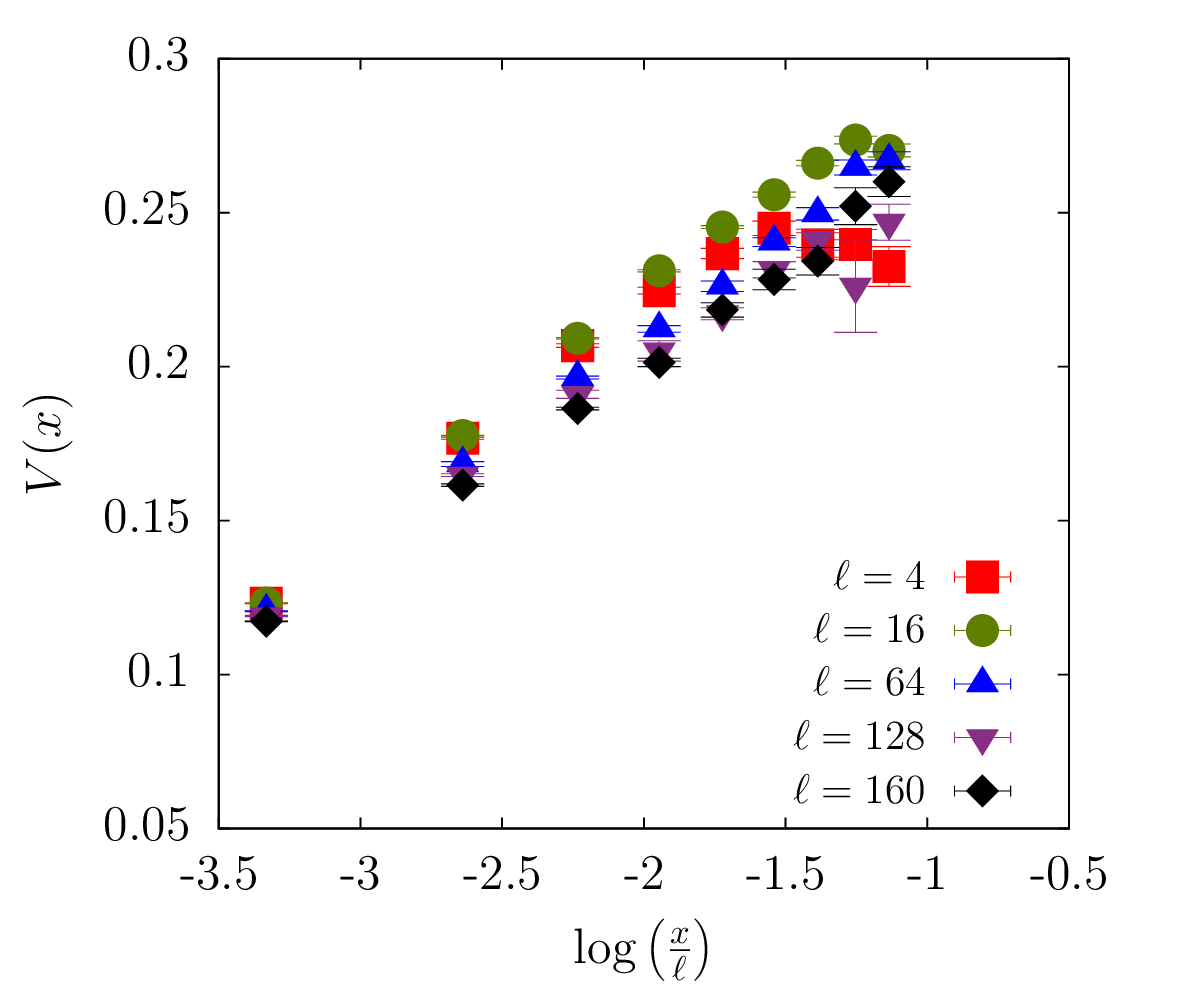}
\end{center}
\caption{
The effective potential, $V(x)$, as a function of the separation, $x$,
 on several different physical volumes as obtained from $x\times t$
 Wilson loops. In the top panel, $V(x)$ is plotted as a function of
$\log(x)$.  The straight lines are the best fits to $V(x)=k \log(x)$. In
the bottom panel, it is plotted as a function of $\log(\frac{x}{\ell})$. 
The data collapse suggests that one cannot set scale using the effective
potential.
}
\eef{wloop}

We start with the results for the effective potential by measuring
the energy, $-\log W(x,t)$, of a  rectangular Wilson loop, $W$, of
size $x\times t$. We regularized the loop by spatially smoothening
the gauge-fields $\theta_1$ and $\theta_2$ perpendicular to $t$,
using 6 levels of APE smearing with the smearing parameter $s=0.5$.
By fitting to
\be
\log(W)=A+V(x)t,
\ee
with $A$ and $V(x)$ as fit parameters, we obtained
the effective potential, $V(x)$, which is shown in the top panel of
\fgn{wloop}. A dominant $\log(x)$ behavior is seen. The deviation from
the $\log(x)$ behavior that is seen at large $x$, diminishes as the
physical volume is increased. A fit of the data to 
\be
V(x) = k \log(x),
\ee
shows that $k$ approaches an infinite volume result. But, $V(x)$
at a fixed $x$, does not seem to have a finite limit. Instead, the
$\log$-term seems to be of the form $\log(\frac{x}{\ell})$, as seen
from the data collapse in the bottom panel of \fgn{wloop}. Therefore,
the effective potential does have a three-dimensional Coulomb-like
behavior but one cannot set a scale using the potential.

\subsection{Scaling of low-lying eigenvalues with physical extent $\ell$}

\bef
\begin{center}
\includegraphics[scale=1.0]{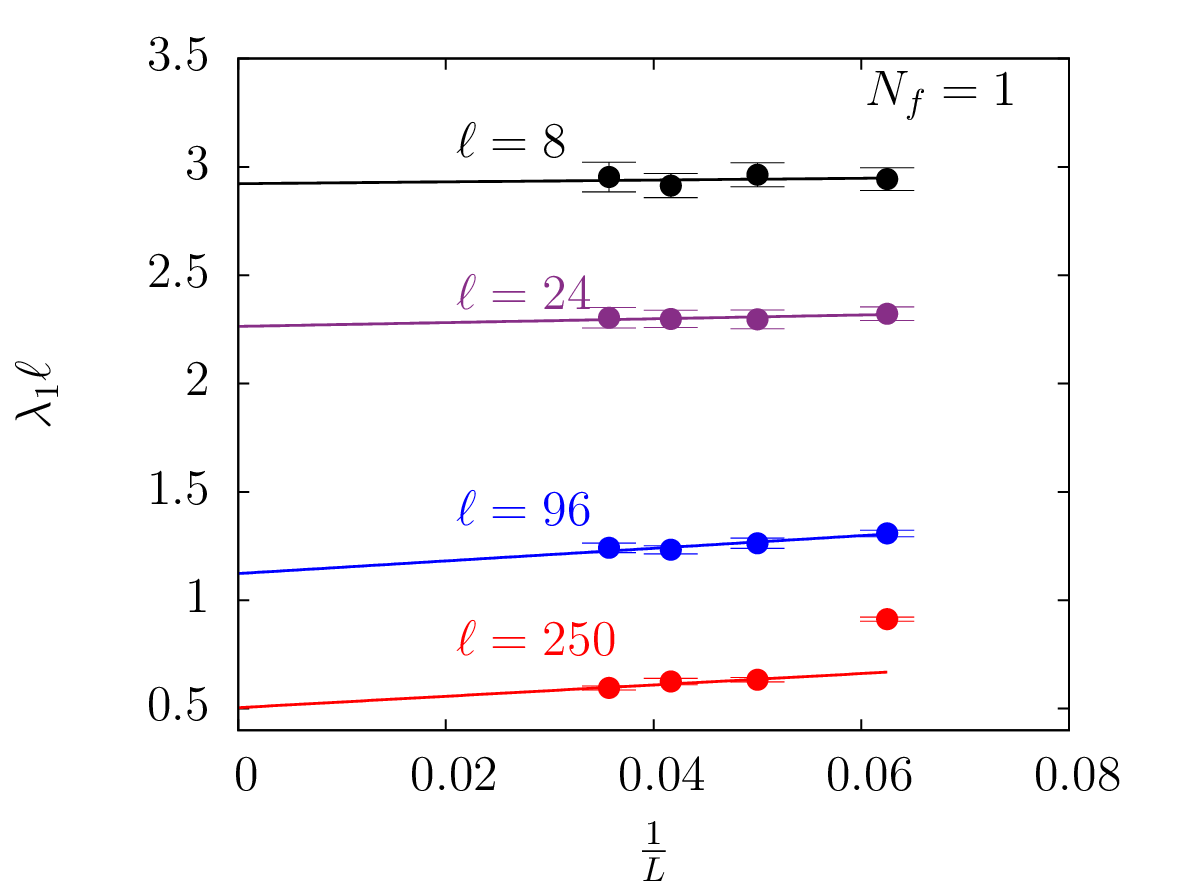}
\end{center}
\caption{
Continuum limit of the lowest eigenvalue at various representative box
sizes $\ell$. The straight lines are the continuum  extrapolations using
a linear $\frac{1}{L}$ fit.
}
\eef{evlimit}

We turn our attention to the 
behavior of the lowest lying eigenvalues as a
function of the box size $\ell$. In \fgn{evlimit}, we show the lattice 
corrections to the smallest dimensionless eigenvalue, $\lambda_1 \ell$,
 at various physical volumes $\ell^3$. As it can be seen, the lattice 
corrections are small at all volumes and under control. Using a linear
$\frac{1}{L}$ extrapolation, we were able to obtain the continuum limit
 of the eigenvalues. Such continuum extrapolations for representative
 volumes are shown by the straight lines in \fgn{evlimit}. We plot
$\lambda_1 \ell$ as a function of $\ell$ for the four largest value of
$L$ in our simulation in the top panel of \fgn{ev1vsl}. In addition we
also show the continuum limit in the same plot.  We expect $\lambda_1
\ell \sim \ell^{-2}$ if this theory has a non-zero bilinear condensate. 
The data shown in the log-log plot is not described by a simple linear
fit.  In the bottom panel of \fgn{ev1vsl}, we compare the continuum
extrapolated $\lambda_1\ell$ with a $\ell^{-2}$ power-law. Large
deviation from the $\ell^{-2}$ behavior is seen at all volumes. To
quantify this statement with some confidence, we fitted the lowest three
eigenvalues to a rational ansatz
\beq
\log\left(\lambda\ell\right)=\frac{a_1-\left(p+\frac{a_2}{\ell}\right)\log(\ell)}{1+\frac{a_3}{\ell}},
\eeq{ansatz}
such that one recovers a power-law as $\ell\to\infty$. Using this,
we extracted the leading power $p$.  The best fit is shown as a solid blue
curve in the bottom panel of \fgn{ev1vsl}.  The $\chi^2/{\rm DOF}$ as a function of $p$
is shown in \fgn{chisq1}. The value of $p$ around 1 seems to be favored
while the value $p=2$, as expected when non-zero bilinear condensate
is present, seems to be ruled out. The value of $\chi^2$ seems to be
minimized around the  same value for all the low-lying eigenvalues,
suggesting that they all scale the same way with $\ell$.

\bef
\begin{center}
\includegraphics[scale=1.0]{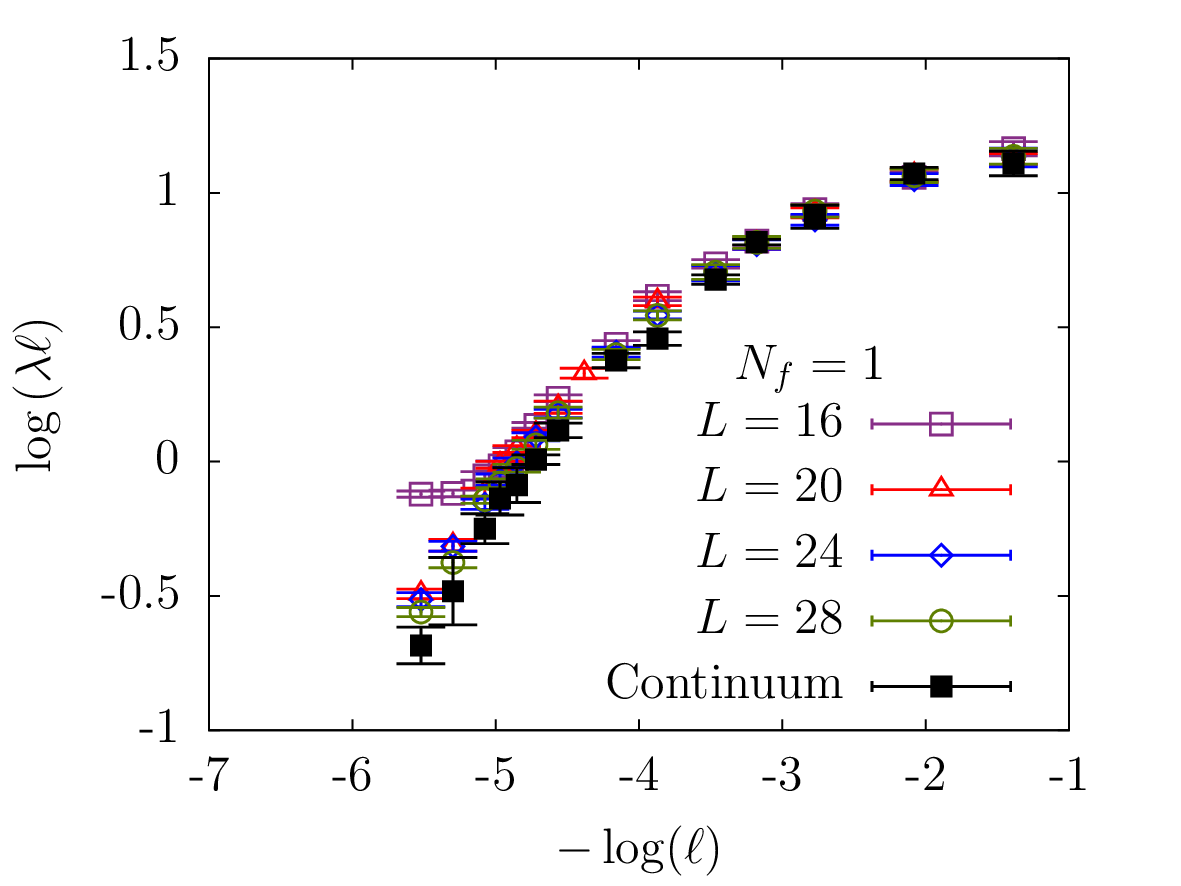}

\includegraphics[scale=1.0]{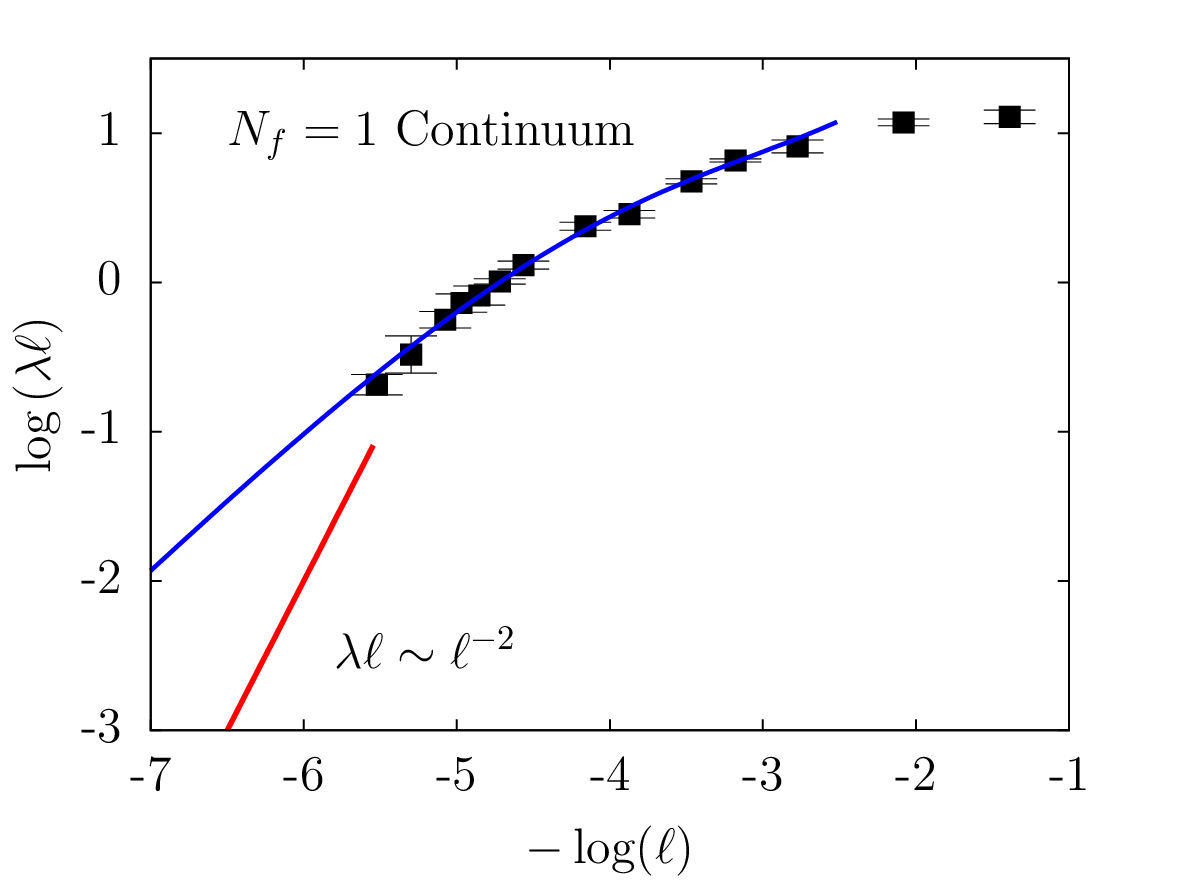}
\end{center}
\caption{
The top panel shows the behavior of the lowest eigenvalue as a function
of the physical volume at finite lattice spacing and in the continuum
limit. The bottom panel compares the $\ell$-dependence of the lowest
eigenvalue (in the continuum limit) with the expected behavior for a
non-zero condensate (red straight line). The solid blue curve is the best fit using
the finite volume ansatz in \eqn{ansatz} with $p=1$.
}
\eef{ev1vsl}

\bef
\begin{center}
\includegraphics[scale=1.0]{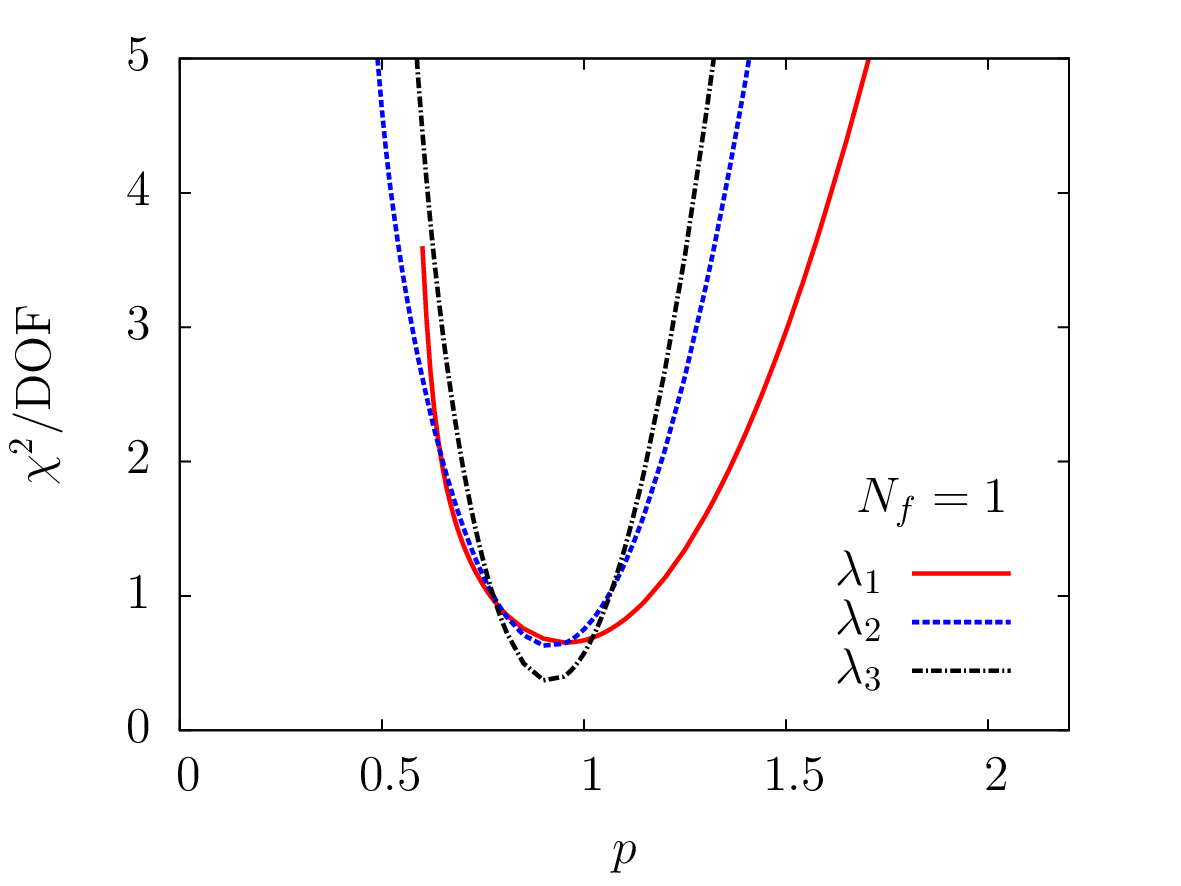}
\end{center}
\caption{
The likelihood of the values of the exponent $p$ describing the asymptotic
behavior $\lambda \ell \sim \ell^{-p}$. The $\chi^2/{\rm DOF}$ for the fit
of the ansatz in \eqn{ansatz} to the $\ell$-dependence of the continuum
extrapolated $\lambda \ell$ is shown as a function of $p$. The degrees
of freedom DOF$=9$ for the fits. The three different curves correspond
to the lowest three eigenvalues. The plot shows that $p=2$, which is
expected when condensate is present, is excluded.
}
\eef{chisq1}

It is possible that our studies have not reached asymptotically
large volumes.  In fact such a possibility has been put forward
in~\cite{Gusynin:2003ww} by studying the effect of an infra-red cutoff
on the gap equation in~\cite{Appelquist:1986qw}. Of course, one cannot
define a condensate at zero momentum in the presence of an infra-red
cutoff but one can ask if the gap equation has a non-trivial solution. It
is argued in~\cite{Gusynin:2003ww} that one need to have $\ell > 200$
to obtain a non-trivial solution for $N_f=1$. This argument is based on
the assumption that the lowest momentum appearing in the fermion loop is
$\pi/\ell$. If instead, we replace the sum over momenta in the fermion
loop by a sum over the eigenvalues of the Dirac operator in the presence
of a gauge field background, a more natural choice for the infra-red
cutoff is the lowest eigenvalue $\lambda_1$ which, in our simulations,
behaves like $1/\ell^2$. Explicitly, at our largest physical extent
$\ell=250$, $\frac{1}{\lambda_1}\approx 500$ which is well inside the
region for a non-trivial solution in~\cite{Gusynin:2003ww}.

\subsection{Comparison to non-chiral random matrix theory}

We can provide further credence to our conclusion of an absence of a
bilinear condensate at $N_f=1$ by borrowing ideas from random matrix
theory. In a theory that has a non-zero bilinear
condensate in three dimensions, we should find low-lying eigenvalues to
scale like $1/\ell^3$ and this behavior should extend for all
eigenvalues below a threshold proportional to $1/\ell^2$. One
expects these low-lying eigenvalues to be  dependent only on the
fluctuations of the zero-mode of the chiral Lagrangian, and hence
determined only by the symmetries  of the low-energy theory. Thus,
a diminishing fraction of eigenvalues ($\ell$ out of the $\ell^3$
eigenvalues), should be described by a random matrix theory (RMT)
which has the same symmetries  as that of the Dirac operator (see
\eqn{diracop}), but the actual number of them would get larger as one
goes to infinite volume. This low-lying spectrum of eigenvalues is
the ergodic regime. The non-chiral RMT which has the same symmetries
as that of QED$_3$ has been  studied in~\cite{Verbaarschot:1994ip}. A
consequence is that the ratios of eigenvalues must be universal and be
described by this  non-chiral RMT. In the top panel of \fgn{ratio},
we plot the histogram $P(\lambda_1/\lambda_2)$ at various volumes. The
expectation from RMT is also shown. A large volume dependence is
seen. In order to take the $\ell\to\infty$ limit, it was convenient 
for us to use the cummulant generating function $G(s)$,
\beq
G(s)=\log\int_0^\infty P(x) e^{-s x} dx\qquad\text{where}\qquad x=\frac{\lambda_1}{\lambda_2}.
\eeq{cumul}
These are shown in the bottom panel of \fgn{ratio}. Using a $[1/1]$
Pad\'e approximant, we extrapolated the $G(s)$ to  its $\ell\to\infty$
limit. This is shown by the green band, labelled $\ell=\infty$, in
\fgn{ratio}. It is clear that there is no agreement with  non-chiral
RMT. Nevertheless, we note that the ratio $\lambda_1/\lambda_2$ does
have a non-trivial limit showing that the different  eigenvalues scale
the same way with volume.  However this scaling is not $1/\ell^3$.
This justifies the inference from \fgn{chisq1}. 

\bef
\begin{center}
\includegraphics[scale=1.0]{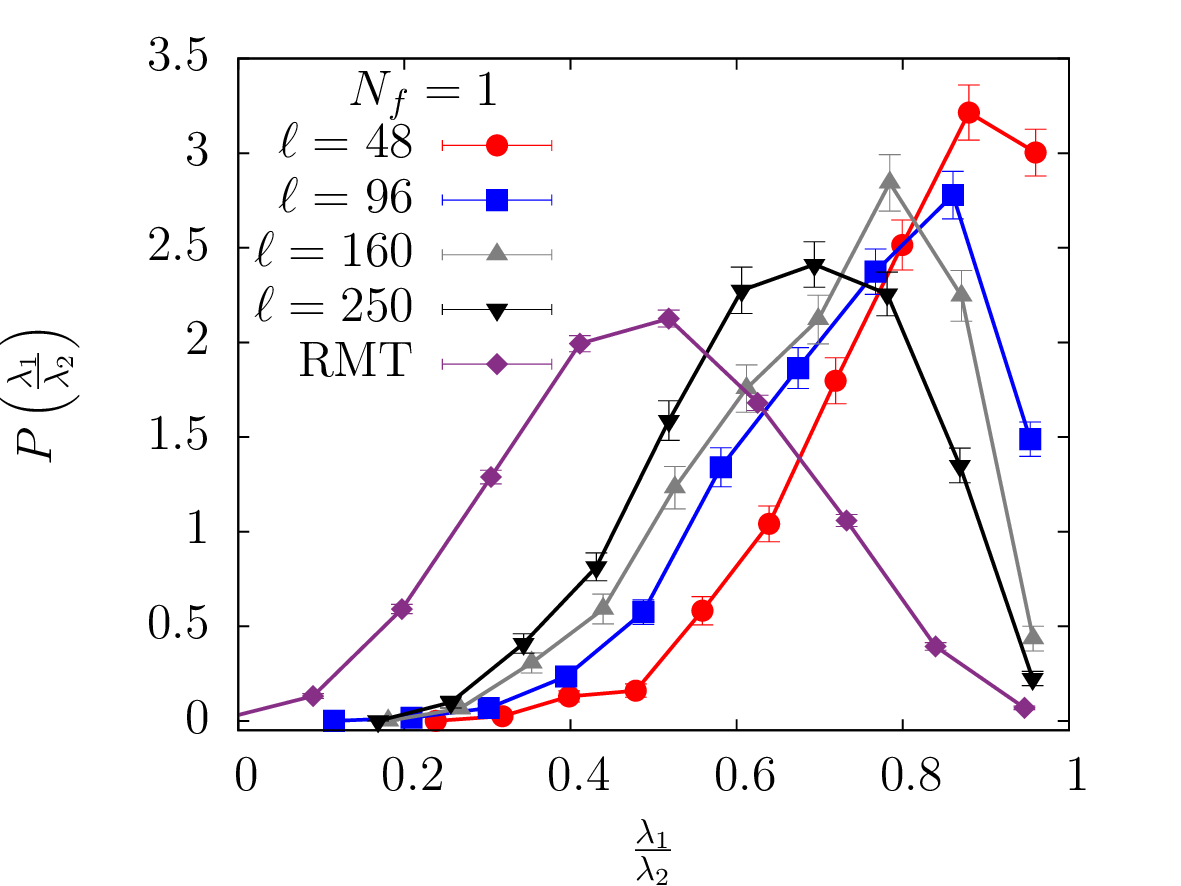}

\includegraphics[scale=1.0]{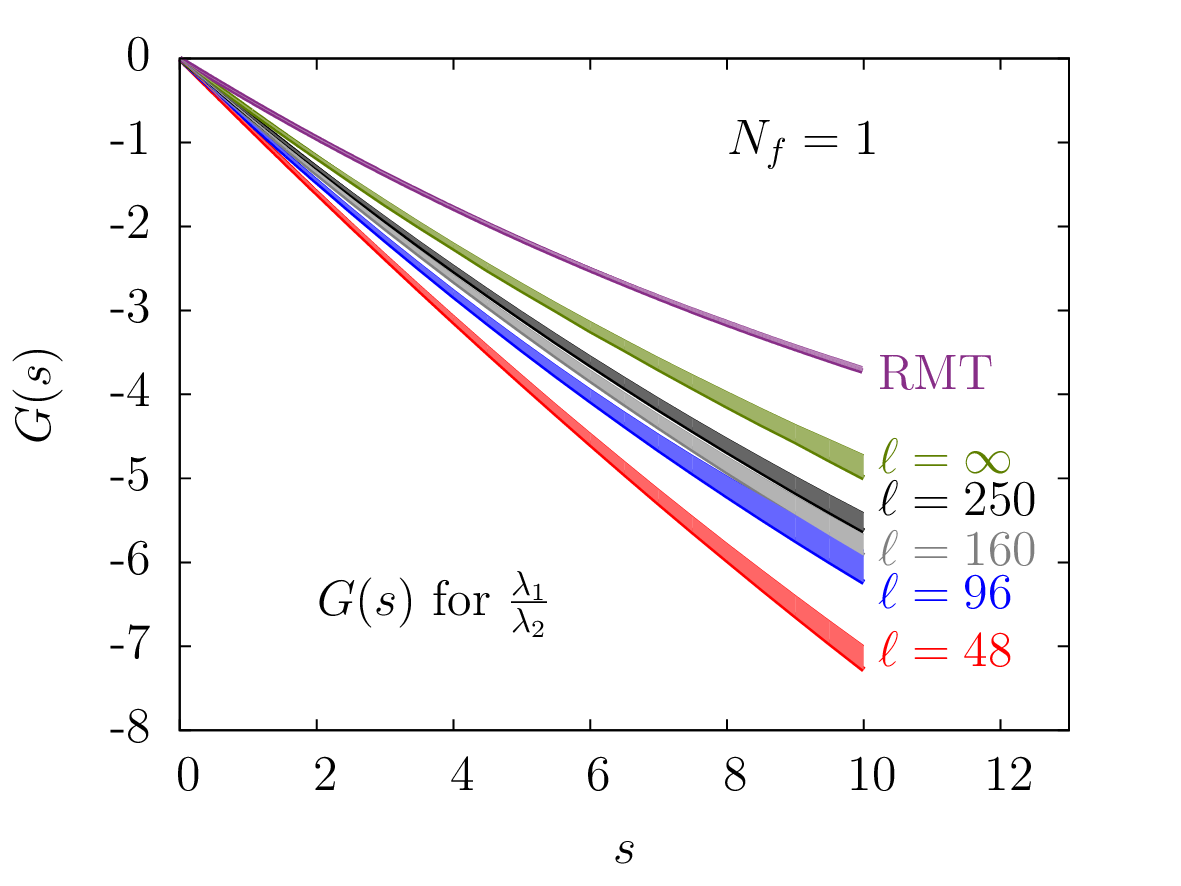}
\end{center}
\caption{
Distribution of $\lambda_1/\lambda_2$. The top panel shows the histogram
of $\lambda_1/\lambda_2$ at various volumes.  The magenta diamonds
correspond to that of non-chiral random matrix theory. The bottom panel
shows the 1-$\sigma$ bands of the cummulant generating function $G(s)$
for the probability distributions in the top panel. The green band
(labelled $\ell=\infty$) is the infinite volume extrapolation of $G(s)$.
The magenta band (labelled as RMT) is the expectation from non-chiral
RMT.  
}
\eef{ratio}

\subsection{Emergence of critical behavior}

In the ergodic regime, the eigenmodes would be delocalized and the
associated Inverse Participation Ratio (IPR),
\be
I_2(\lambda) = \int \left(\psi_\lambda^*(x)\psi_\lambda(x)\right)^2 d^3 x\qquad\text{with normalization}\qquad
 \int \psi_\lambda^*(x)\psi_\lambda(x) d^3 x =1,
\ee
would scale as $1/\ell^3$. The IPR for the lowest mode is plotted
in \fgn{ipr1}. We see a significant deviation away from the ergodic
behavior. Instead, the modes seem to be delocalized, but multi-fractal
\ie,
\be
I_2(\ell) \sim \ell^{-3+\eta}\qquad\text{with}\qquad \eta\ne 0.\label{chiscale}
\ee
In our case, the value of $\eta$ is 0.32(1). This multi-fractal scaling 
suggests a critical behavior, which is further quantified below. It is usual to draw analogy
between the broken phase where an RMT description is possible  and a
metallic state~\cite{Osborn:1998nf}.  In this spirit, this behavior
of IPR is reminiscent of the behavior of electron wavefunction at a
metal-insulator critical point.

\bef
\begin{center}
\includegraphics[scale=1.0]{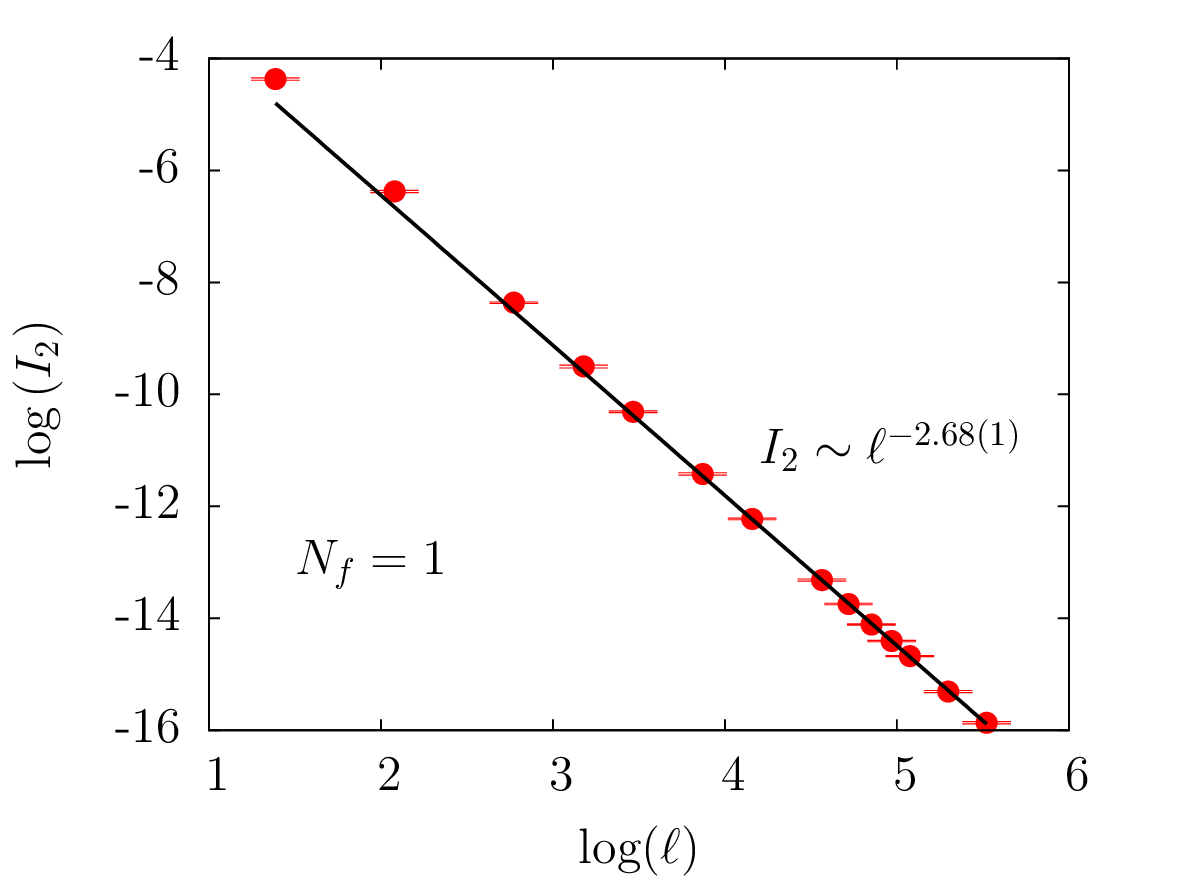}
\end{center}
\caption{
A plot of the IPR of the lowest eigenmode as a function of the physical
volume. The red circles are the data from our simulation.  The black
line is the power-law behaviour ( $I_2\sim \ell^{-2.68}$) seen at large
enough box sizes $\ell$. The ergodic behaviour would have been $I_2\sim
\ell^{-3}$.
}
\eef{ipr1}

\bef
\begin{center}
\includegraphics[scale=1.0]{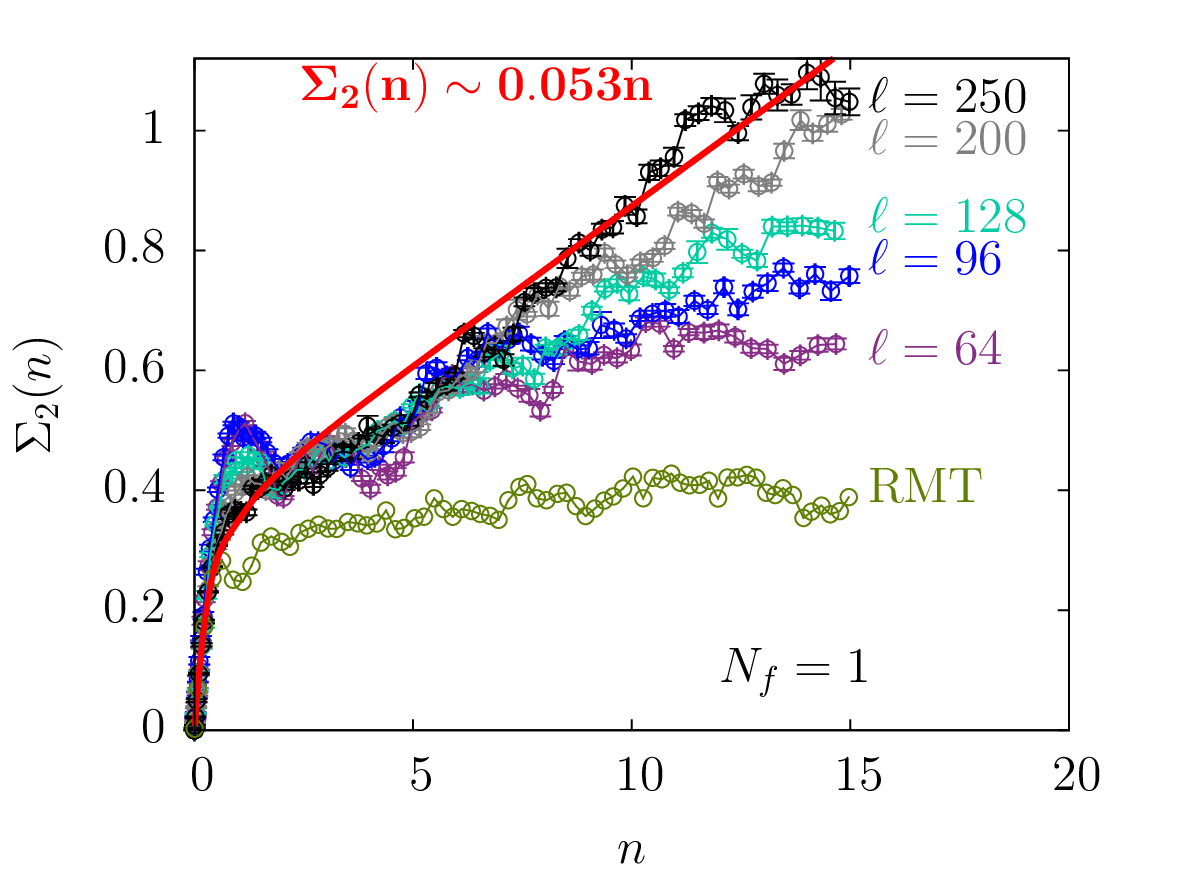}
\end{center}
\caption{
A plot of the number variance $\Sigma_2(n)$ as a function of $n$,
the number of eigenvalues below a certain scale,
 for different physical volumes. The green points (labelled RMT)
are the ones expected from non-chiral random matrix theory, which
is ergodic. The red solid curve is obtained from the $q$-Hermite
random matrix model (refer \eqn{qherm}), which is critical and has
an asymptotic behavior $\Sigma_2(n)\sim\chi n$.  The value of $\chi$
was set to $\eta/6=0.053$ as inferred from the $\ell$-dependence
of IPR in \fgn{ipr1}.
}
\eef{numvar1}

We proceed to compare the behavior of the IPR of the lowest mode
with that of number variance, a quantity that does not depend
on the microscopic spectral density, but is a measure of ergodic
behavior~\cite{Osborn:1998nf,Altschuler:1986zh,Altschuler:1988al,Chalker:1996kr}.
 The number variance is computed as follows: having chosen a
$\lambda$, we find the number of eigenvalues $n$ below that scale per
configuration. Then, the number variance is 
\be
\Sigma_2(n)={\rm Var}(n).
\ee
Normally, one wants to study this quantity in the bulk away from the
edge such that critical behavior sets in at the transition between the
ergodic,
\be
\Sigma_2(n) \sim \ln n,
\ee
and diffusive,
\be
\Sigma_2(n) \sim n^{\frac{3}{2}},
\ee
behavior. As noted earlier, the region showing the ergodic behavior
should increase linearly with $\ell$. In \fgn{numvar1}, we show the
behavior of $\Sigma_2(n)$ for several physical volumes.  The behavior
at large volumes show a linear rise over a wide range with no
region that shows ergodic behavior (labelled as RMT).  Furthermore,
the slope of the linear growth, 
\be
\Sigma_2(n) = \chi n,
\ee
is $\chi=0.057$, which is 
consistent with the critical relation~\cite{Chalker:1996kr}
\be
\eta=6\chi,
\ee 
between the slope of the number variance and the multi-fractal
dimension $\eta$  of the lowest  mode. This critical behavior seen
right from the lowest mode is surprising. There are random matrix
theory models with special choices of potentials  where all states
exhibit critical behavior~\cite{Kravtsov:1997zz}. An exactly solved
model in this critical class of random matrix theories is the
$q$-Hermite model introduced in~\cite{Muttalib:1993zz}, and it has
a single tunable parameter, which is the slope $\chi$ at large $n$.
The number variance in the model is given
by~\cite{blecken1994transitions,muttalibcomm}
\be
\Sigma_2(n)= n - 2\pi^2\chi^2\int_0^n (n-\xi)\left[\frac{\sin(\pi\xi)}{\sinh(\pi^2\chi\xi)}\right]^2 d\xi,
\label{qherm}
\ee
which behaves asymptotically as $\Sigma_2(n)\sim\chi n$.  We compare
our data to that of the $q$-Hermite model as follows: using the
value $\eta=0.32$ that we determined from the IPR of the lowest
mode (refer \fgn{ipr1}), we calculated the number variance for the
$q$-Hermite model using the above equation with $\chi=\eta/6$.  This
is shown as the red solid curve in \fgn{numvar1}. An agreement is
seen, even for small $n$ (with no fitting involved).

\bef
\begin{center}
\includegraphics[scale=0.9]{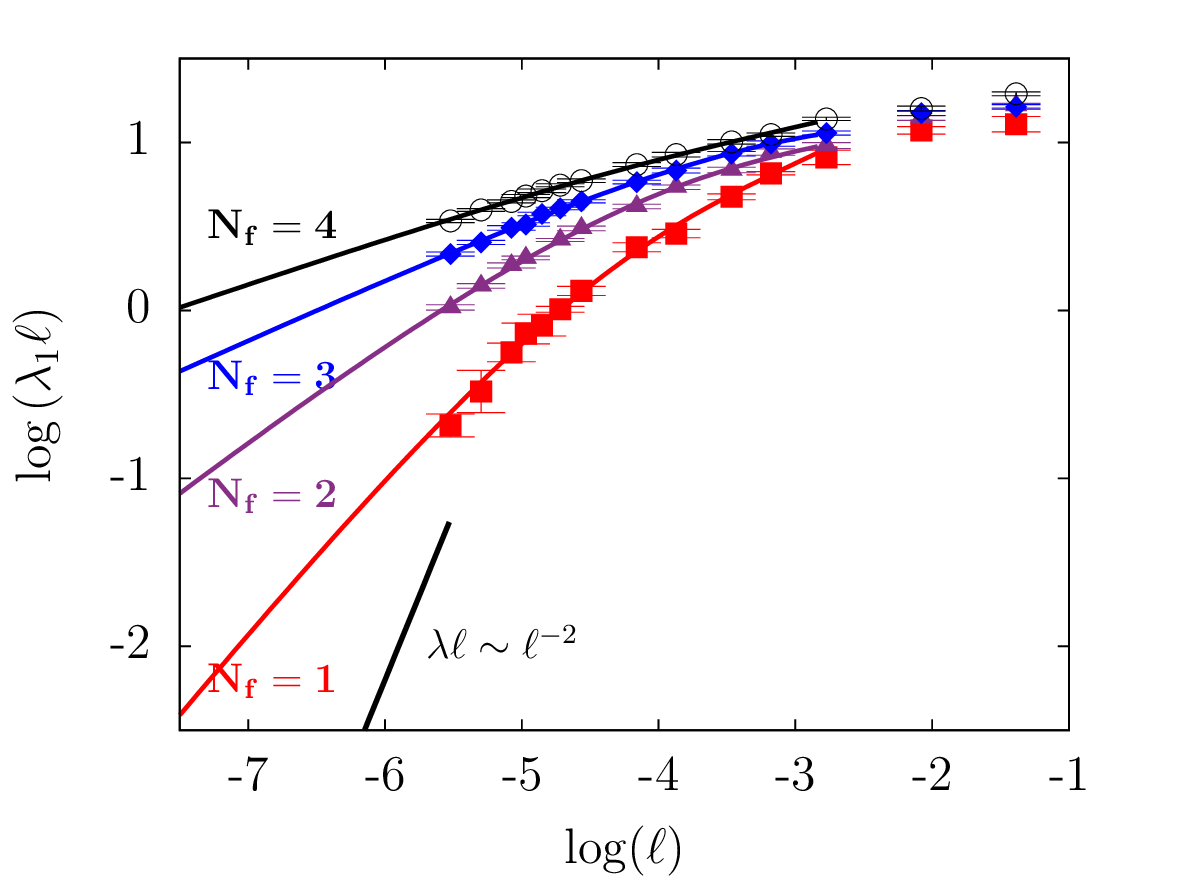}

\includegraphics[scale=0.9]{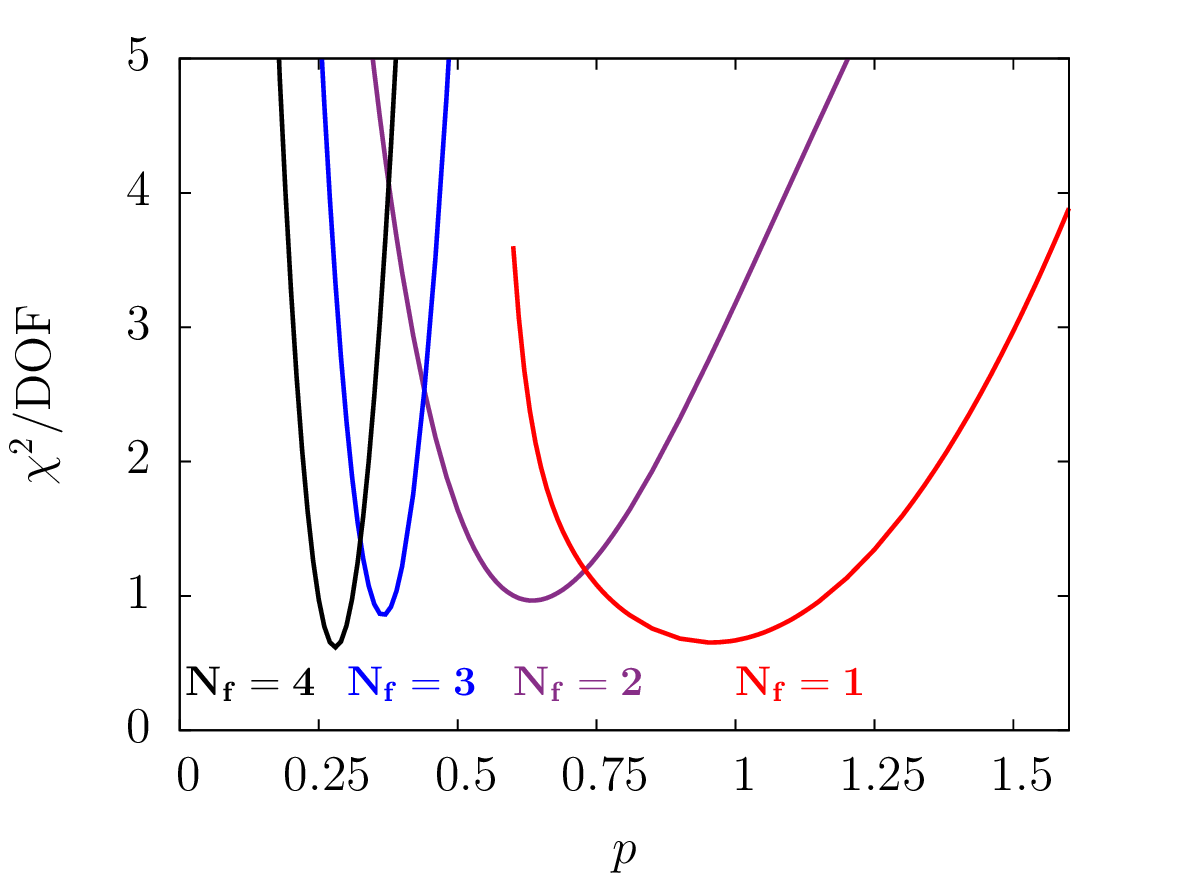}
\end{center}
\caption{
The top panel shows the $\ell$-dependence of the continuum extrapolated
dimensionless eigenvalue, $\lambda_1\ell$, for $N_f=1,2,3,4$ (as
labelled on the left of the plot). The power-law $\ell^{-2}$, expected
if a condensate is present, is shown for comparison. The different
curves are the best fits of the finite volume ansatz in \eqn{ansatz}
to the data. The bottom panel shows the $\chi^2/{\rm DOF}$ for the fit
of the finite volume ansatz to the data, as a function of exponent $p$,
which describes the asymptotic behavior $\lambda_1\ell\sim\ell^{-p}$. The
most likely value of $p$ seems to decrease monotonically with $N_f$.  
}
\eef{evvsl}

\section{Results for $N_f > 1$}\label{sec:resng1}

\bef
\begin{center}
\includegraphics[scale=0.70]{IPR_Nf1.ps}
\includegraphics[scale=0.70]{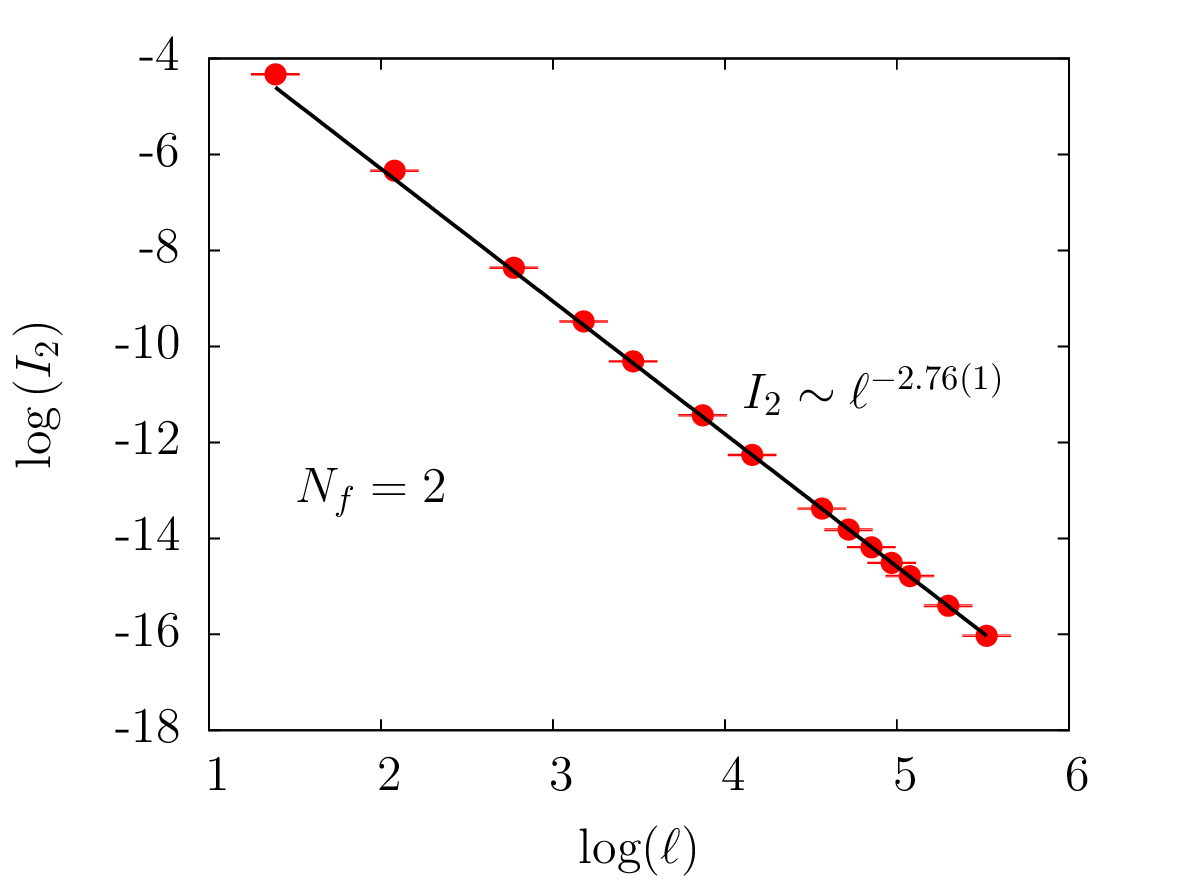}
\includegraphics[scale=0.70]{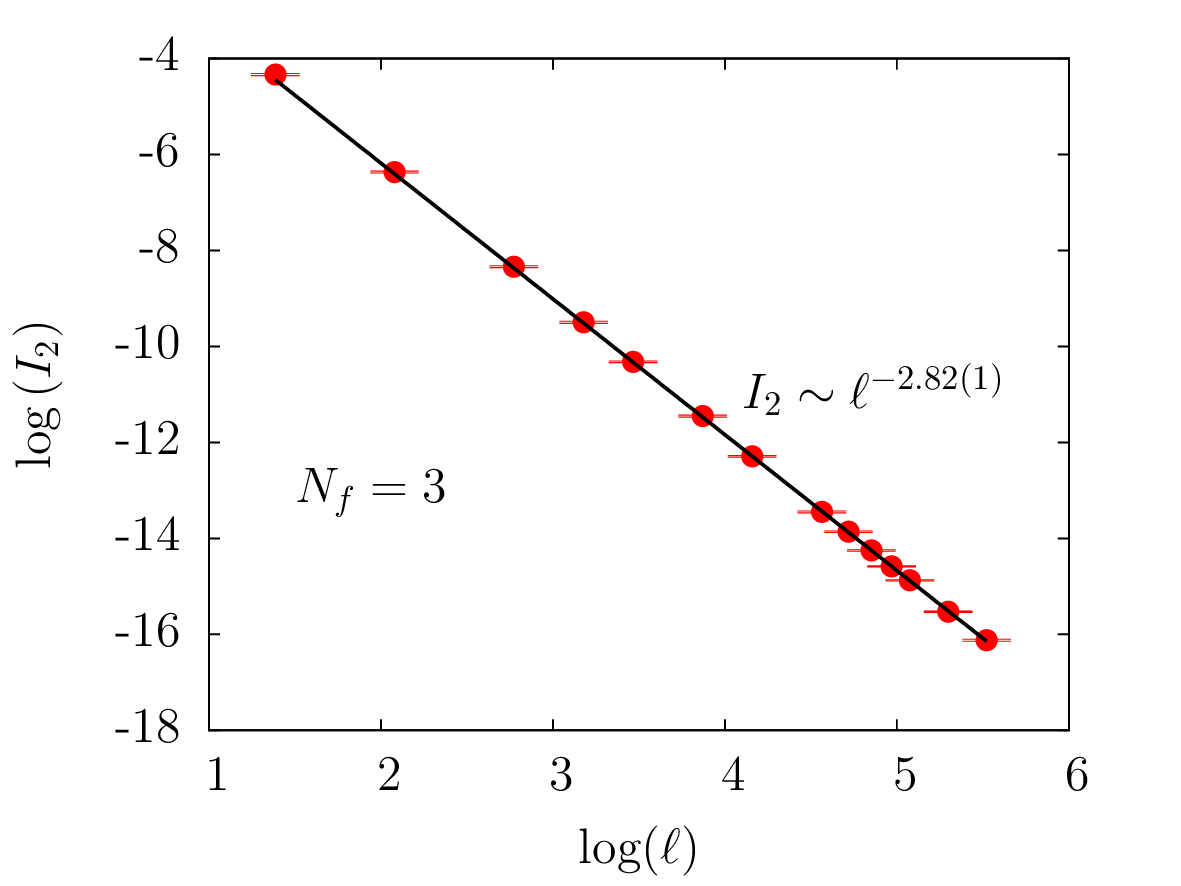}
\includegraphics[scale=0.70]{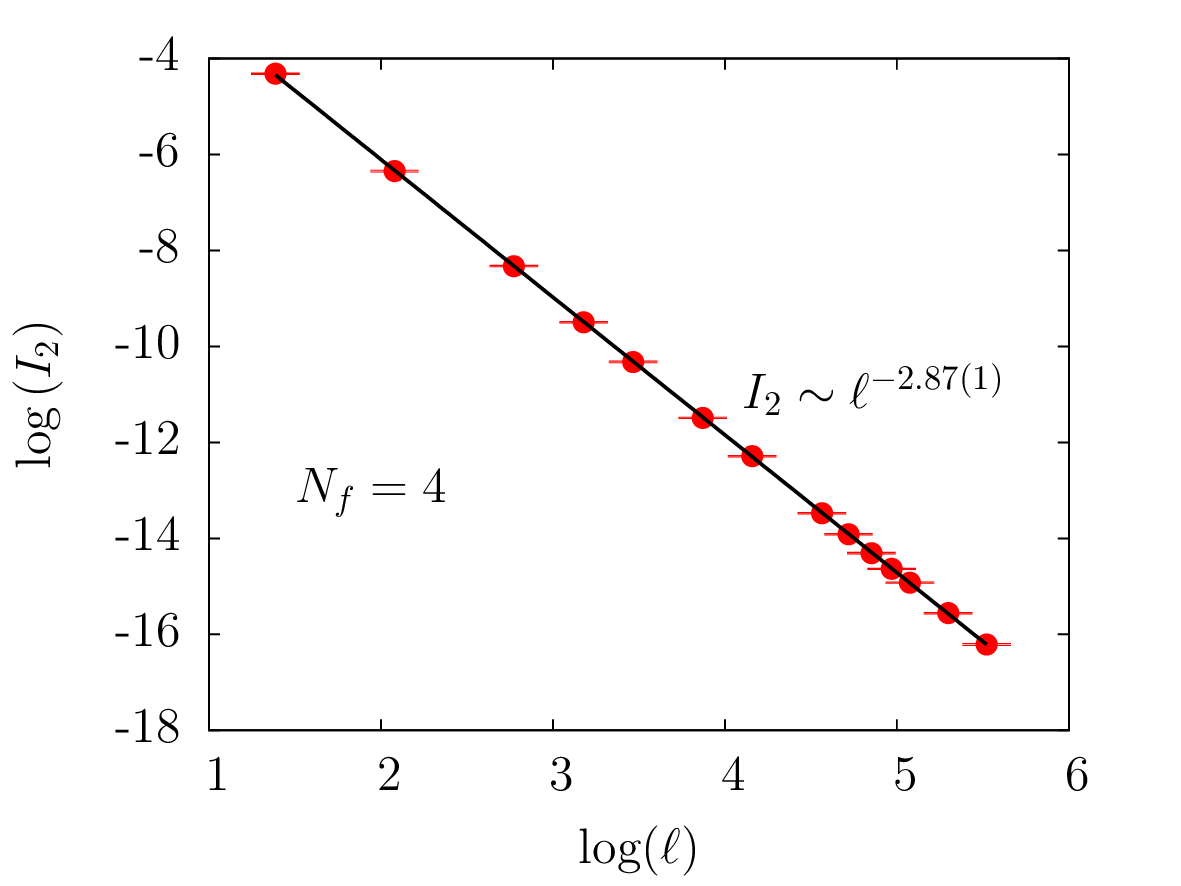}
\end{center}
\caption{
A plot of the IPR, $I_2$, of the lowest eigenmode as a function of the
physical extent of the box, $\ell$, for $N_f=1$ (top left), $N_f=2$
(top right), $N_f=3$ (bottom left) and $N_f=4$ (bottom right). The black
lines are the best fits of the power law, $I_2\sim \ell^{-3+\eta}$,
to the IPR at large volumes.
}
\eef{iprn}

Our results for $N_f>1$ are provided with the aim of comparing them
to the $N_f=1$ results. With that in mind, some of the previous plots
are repeated for convenience to the reader. The behavior of the lowest
eigenvalue as a function of the physical extent of the box, $\ell$,
is shown in the top panel of \fgn{evvsl}.  Clearly, at any $\ell$, the
value of the eigenvalue itself increases with $N_f$, and this is due
to the suppression of the low eigenvalues by the fermion determinant.
In addition, all the four cases show a behavior that is consistent
with an absence of a bilinear condensate.  The $\chi^2$-values for the
different values of the exponent $p$ which describes the asymptotic
behavior $\lambda_1\ell\sim\ell^{-p}$ (refer to \eqn{ansatz}), is
shown in the bottom panel of \fgn{evvsl}.  From this, we estimate the
values of $p$ to be $0.97^{+0.34}_{-0.30}$, $0.63^{+0.22}_{-0.15}$,
$0.37^{+0.05}_{-0.06}$ and $0.28^{+0.05}_{-0.06}$ for $N_f=1,2,3$ and
4 respectively. The values of $p$ seem to monotonically decrease with
$N_f$. In fact, the trend seems to be consistent with $p=\frac{1}{N_f}$
behavior.

\bef
\begin{center}
\includegraphics[scale=0.70]{nvar_nf1.ps}
\includegraphics[scale=0.70]{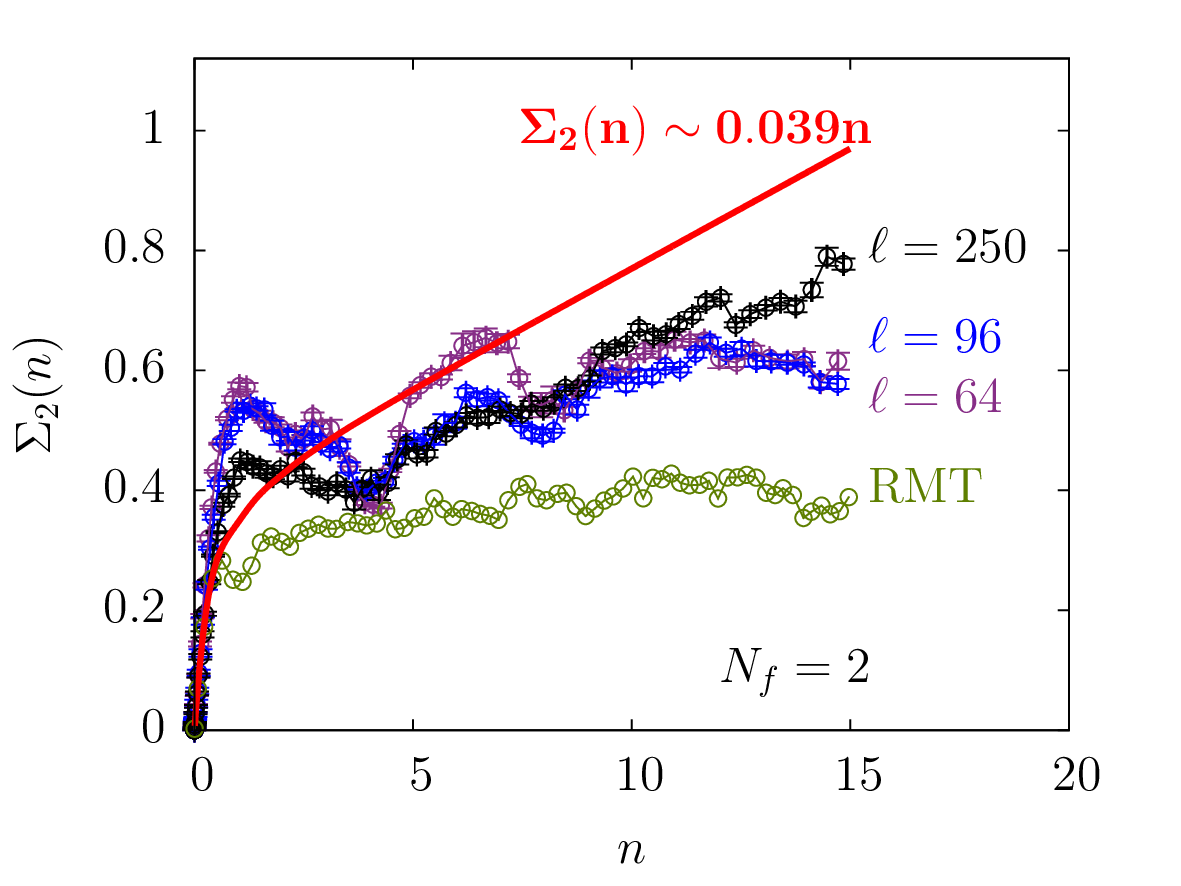}
\includegraphics[scale=0.70]{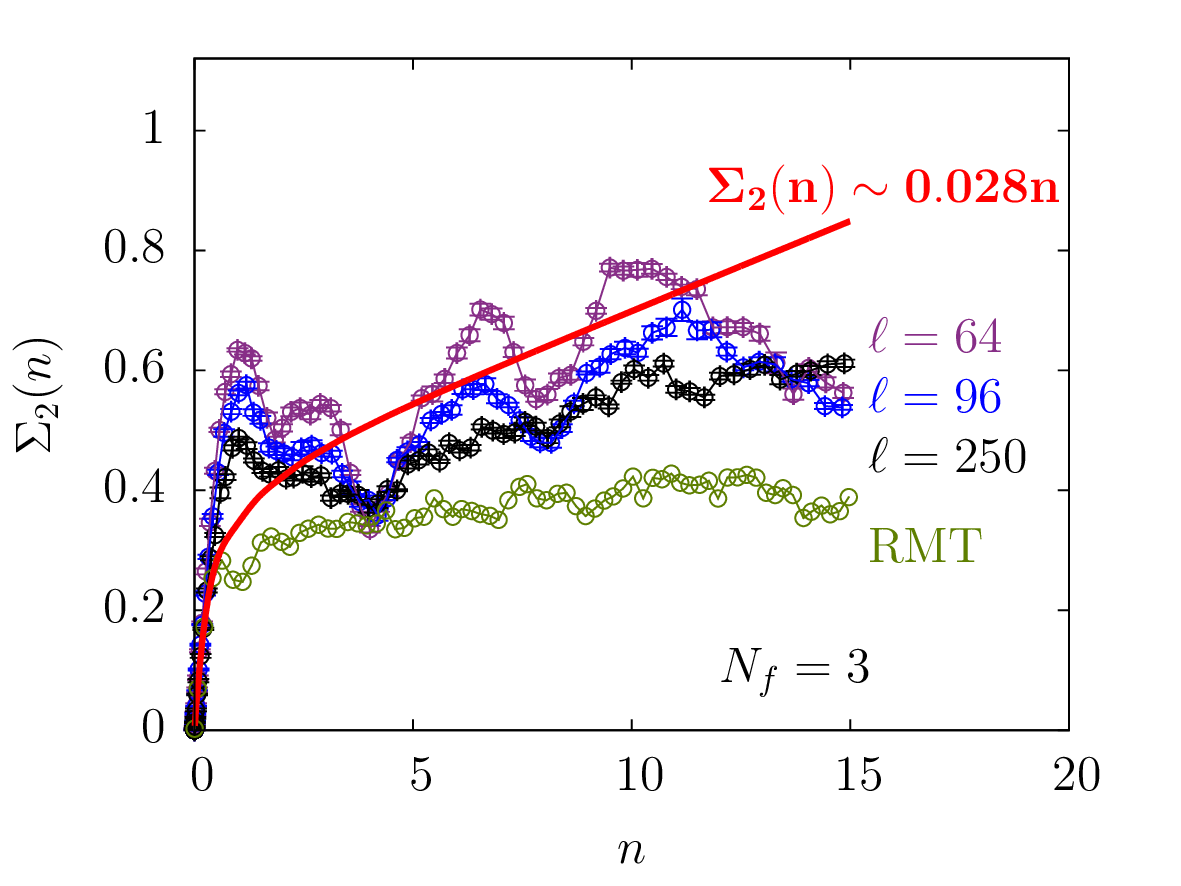}
\includegraphics[scale=0.70]{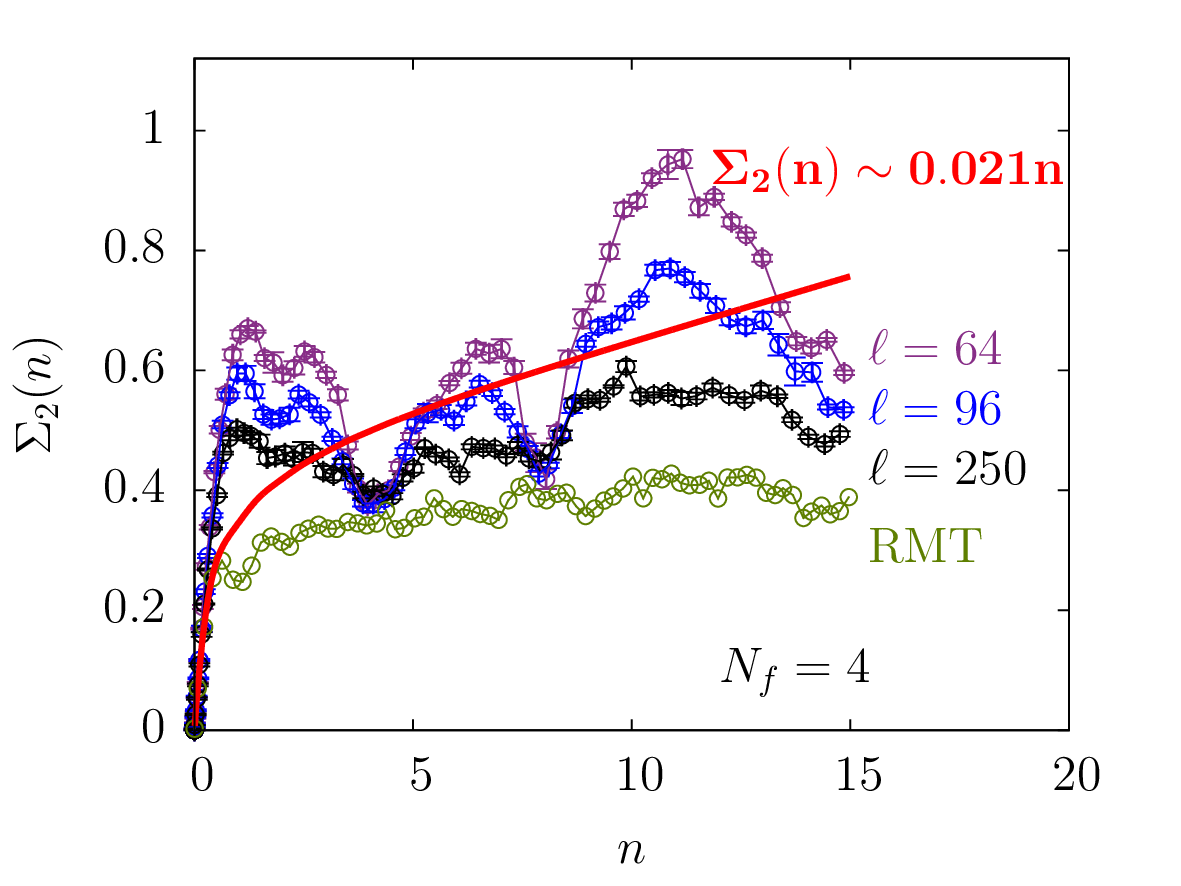}
\end{center}
\caption{
A plot of the number variance for several different physical volumes for
$N_f=1$ (top left), $N_f=2$ (top right), $N_f=3$ (bottom left) and $N_f=4$
(bottom right). The red solid curves are obtained from 
the $q$-Hermite random matrix model (refer \eqn{qherm}), tuned to have 
a behavior $\Sigma_2(n)\sim\frac{\eta}{6}n$,  at large $n$, with 
$\eta$ determined from the IPR of the lowest mode (refer \fgn{iprn}).
}
\eef{numvarn}

In \fgn{iprn}, we compare the IPR of the lowest mode for $N_f=1$
(top left panel) to that of  $N_f=2$ (top right), $N_f=3$ (bottom
left) and $N_f=4$ (bottom right). The black lines are the power-law
fits, $\ell^{-3+\eta}$, to the data at large volumes. In all the
four cases, we see systematic deviation away from the ergodic
$\ell^{-3}$ behavior; instead, we see a multi-fractal scaling at
all $N_f$.  The deviation from the ergodic behavior gets smaller
as we increase $N_f$. In \fgn{numvarn}, we compare the number
variance  of $N_f=1$ (top left panel) with that of $N_f=2$ (top
right), $N_f=3$ (bottom left) and $N_f=4$ (bottom right) for different
choices of physical volume.  Let us first focus on the $N_f=2$ case.
The signature of critical behavior is the linear rise of $\Sigma_2(n)$.
The linear rise seen for $N_f=1$ is also present for $N_f=2$.  As
explained in \scn{resn1}, the number variance for the critical
$q$-Hermite random matrix model, tuned to have an asymptotic behavior
$\Sigma_2(n)\sim\frac{\eta}{6} n$, is also shown as a red solid
curve.  At the volumes we simulated, an agreement is not seen for
$N_f=2$.  Nevertheless, the linear rise in $\Sigma_2(n)$ approaches
$\eta/6$ at large volumes (as seen by comparing the slope of the
linear rise in the data to that of the red solid curve).  The trend
in the data is for the linear segment of $\Sigma_2(n)$ to shift
upwards towards the critical random matrix model with increasing
volume.  To check if an agreement is seen at even larger physical
volumes, requires lattices with larger $L$ in order to control lattice
artifacts. Such a computation is beyond the scope of this work.
The presence of a possible linear rise for $N_f=3$ and $N_f=4$ is
marred by an oscillatory behavior, perhaps because of the comparatively
small value of slope as inferred from $\eta$.

\section{Conclusions}\label{sec:conc}

We have numerically studied a parity invariant formulation of QED
in three dimensions using Wilson fermions. We investigated theories
with $2,4,6$ and $8$ flavors of massless two component fermions.
We used the behavior of low-lying eigenvalues of the four component
Dirac operator to investigate the presence or absence of a bilinear
condensate that preserves parity. Our computations were performed
on several physical volumes and lattice spacings.  The resulting
low-lying spectrum did not exhibit a $\ell^{-3}$ dependence on the
physical linear extent, $\ell$, of the three-dimensional symmetric
periodic box for any of the theories studied here. A study of the
inverse participation ratio of the eigenvectors associated with the
low-lying eigenvalues shows that the modes exhibit critical behavior
when the scaling exponent was compared to the linear rise of the
number variance associated with the low-lying eigenvalues. Furthermore,
the agreement of the number variance for $N_f=1$ between a $q$-Hermite
random matrix model and our data warrants further comparison with
a critical random matrix model.  For example, we plan to compare
the distribution of low-lying eigenvalues of QED$_3$ with that of
a critical random matrix model.  We also plan to study non-abelian
fields coupled to massless fermions. These theories are expected
to be different since there is a non-vanishing string tension in
pure gauge theories.

\acknowledgments

We would like to thank Shinsuke Nishigaki, Jac Verbaarschot and Rohana
Wijewardhana for extensive discussions.  R.N. would like to additionally
thank Jac Verbaarschot for the invitation to the Simons Center Workshop
on Random Matrix Theory, Integrable Systems, and Topology in Physics
which resulted in useful discussions with Gernot Akemann, Takuya
Kanzawa, Tamas Kovacs and Ismail Zahed. We would like to particularly
thank Takuya Kanzawa for bringing~\cite{Elias:2011gp} to our attention
where the authors experimentally demonstrate that there is no gap in
suspended graphene, a system where the low-lying modes are expected to be
described by three-dimensional relativistic QED~\cite{Miransky:2015ava}.
We would like to thank Khandker A.~Muttalib for facilitating
a quantitive comparison of our data with the $q$-Hermite random matrix model.
All computations in this paper were made on the JLAB computing clusters
under a class B project.  The authors acknowledge partial support by
the NSF under grant number PHY-1205396 and PHY-1515446.

\appendix
\section{Details of the two component Sheikhoslami-Wohlert-Wilson-Dirac operator}
\label{sec:swwd}
We are dealing with abelian gauge theory, therefore we smeared the
gauge-fields $\theta_i$ and then constructed the smeared links from
them. Let the directions $\hat j$ and $\hat k$ be orthogonal to $\hat
i$. Explicitly, the HYP smeared field $\theta^s_i(n)$ is given by 
\bea
\theta^s_i(n)&=&s_2^2 \Bigg[\frac{1}{4} \theta _{i }(n-\hat j -\hat k )+\frac{1}{4} \theta _{i }(n+\hat j -\hat k )+\frac{1}{4} \theta _{i }(n-\hat j +\hat k )+\frac{1}{4}
   \theta _{i }(n+\hat j +\hat k )\cr
& &-\frac{1}{8} \theta _{j }(n+\hat i -\hat k )+\frac{1}{8} \theta _{j }(n+\hat i -\hat j -\hat k )-\frac{1}{8} \theta _{j }(n+\hat i
   +\hat k )+\frac{1}{8} \theta _{j }(n+\hat i -\hat j +\hat k )\cr
& &-\frac{1}{8} \theta _{k }(n+\hat i -\hat j )-\frac{1}{8} \theta _{k }(n+\hat i +\hat j )+\frac{1}{8}
   \theta _{k }(n+\hat i -\hat j -\hat k )+\frac{1}{8} \theta _{k }(n+\hat i +\hat j -\hat k )\cr
& &-\frac{1}{4} \theta _{i }(n-\hat j )-\frac{1}{4} \theta _{i }(n+\hat j
   )-\frac{1}{4} \theta _{i }(n-\hat k )-\frac{1}{4} \theta _{i }(n+\hat k )\cr
& &+\frac{1}{8} \theta _{j }(n-\hat k )-\frac{1}{8} \theta _{j }(n-\hat j -\hat k
   )+\frac{1}{8} \theta _{j }(n+\hat k )-\frac{1}{8} \theta _{j }(n-\hat j +\hat k )\cr
& &+\frac{1}{8} \theta _{k }(n-\hat j )+\frac{1}{8} \theta _{k }(n+\hat j
   )-\frac{1}{8} \theta _{k }(n-\hat j -\hat k )-\frac{1}{8} \theta _{k }(n+\hat j -\hat k )\Bigg]\cr
& &+s_1  \Bigg[\frac{1}{4} \theta _{i }(n-\hat j
   )+\frac{1}{4} \theta _{i }(n+\hat j )-\frac{1}{4} \theta _{j }(n+\hat i )+\frac{1}{4} \theta _{j }(n+\hat i -\hat j )\cr
& &+\frac{1}{4} \theta _{i }(n-\hat k
   )+\frac{1}{4} \theta _{i }(n+\hat k )-\frac{1}{4} \theta _{k }(n+\hat i )+\frac{1}{4} \theta _{k }(n+\hat i -\hat k )-\theta _{i }(n)\cr
& &+\frac{\theta _{j
   }(n)}{4}-\frac{1}{4} \theta _{j }(n-\hat j )+\frac{\theta _{k }(n)}{4}-\frac{1}{4} \theta _{k }(n-\hat k )\Bigg]+\theta _{i }(n).
\eea
We set the smearing parameters to $s_1=0.6$ and $s_2=0.5$.
The unsmeared and smeared link variables are given by
\be 
U_i(n) = e^{i\theta_i(n)}\qquad\text{and}\qquad V_i(n) = e^{i\theta^s_i(n)},
\ee
respectively.
The Sheikhoslami-Wohlert-Wilson-Dirac operator for the two-component fermion, in lattice units,  is
\bea
C_W(n,m)&=&(-3+M_P)\delta_{n,m}+\frac{1}{2}\sum_{i=1}^3\left\{\left(1+\sigma_i\right)V_i(n)\delta_{n+i,m}+\left(1-\sigma_i\right)V^*_i(n-\hat i)\delta_{n-i,m}\right\}\cr
&&+i\frac{\kappa_{\scriptscriptstyle SW}}{8} \delta_{n,m} \sum_{i,j,k=1}^3 \epsilon_{ijk}  C_{ij}(n) \sigma_k,
\eea
where the clover term is
\be
C_{ij}(n) = P_{ij}(n) + P_{ij}(n-\hat i) +  P_{ij}(n-\hat j) + P_{ij}(n-\hat i-\hat j),
\ee
in terms of the plaquette,
\be
P_{ij}(n)=U_i(n)U_j(n+\hat i)U^*_i(n+\hat j)U^*_j(n).
\ee
We used the value $\kappa_{\scriptscriptstyle SW} = 0.5$ at all $\ell$
and $L$. Using these values of $\kappa_{\scriptscriptstyle SW}$ and
the smearing parameters, $s_1$ and $s_2$, we found the additive mass
renormalization to be greatly reduced even in our coarsest lattices.

\section{Simulation details}\label{sec:simpar}

\bef
\begin{center}
\includegraphics[scale=0.90]{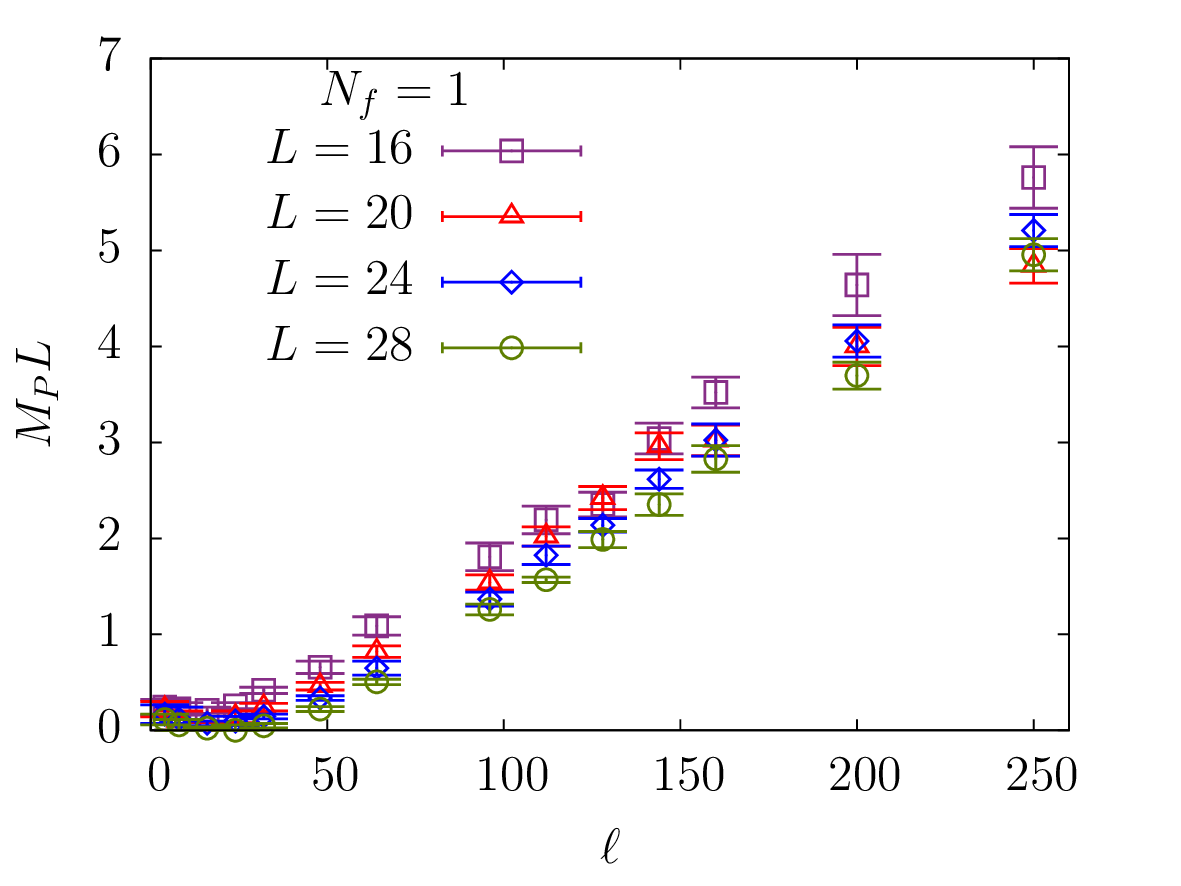}
\end{center}
\caption{
The tuned mass in physical units, $M_P L$, is shown as a function of $\ell$ at different lattice sizes $L$.
}

\eef{mw}

\bet
\begin{center}
\begin{tabular}{|c||c|c|c|c|}
\hline
$\ell$ & \multicolumn{4}{|c|}{$M_P$} \\
\cline{2-5}
 & $L=16$ & $L=20$ & $L=24$ & $L=28$ \\
\hline
4 & 0.015(5) & 0.011(4) & 0.007(4) & 0.006(3) \\
8 & 0.014(4) & 0.005(5) & 0.006(2) & 0.002(2)\\
16 & 0.013(2) & 0.003(1) & 0.003(1) & 0.0008(4) \\
24 & 0.016(2) & 0.007(3) & 0.004(2) & 0.000(2)\\ 
32 & 0.026(2) & 0.012(2)& 0.006(1) & 0.0017(8)\\
48 & 0.041(4) & 0.023(2)& 0.014(1) & 0.0079(9)\\
64 & 0.068(6) & 0.041(3)& 0.027(3) & 0.018(1)\\
96 & 0.113(9) & 0.077(4)& 0.057(3) & 0.045(2)\\
112 & 0.137(9) & 0.101(5)&0.076(4) &  0.056(1) \\
128 & 0.147(8) & 0.121(6)& 0.089(3) & 0.071(3) \\
144 & 0.19(1) & 0.148(7) & 0.109(4) &0.084(4)\\
160 & 0.22(1) & 0.151(8) & 0.126(7) &0.101(5)\\
200 & 0.29(2) & 0.20(1) & 0.169(7) & 0.132(5) \\
250 & 0.36(2) & 0.242(9) & 0.217(7) & 0.177(6) \\
\hline
\end{tabular}
\end{center}
\caption{Simulation parameters for $N_f=1$.}
\eet{nf1data}

The free parameters are the physical extent of the box, $\ell$, the
lattice size, $L$, the tuned Wilson mass, $M_P$, and the number 
of flavors, $N_f$. We used $N_f=$1, 2, 3 and 4 in our simulations. 
The continuum limit is taken by taking $L\to\infty$ keeping $\ell$
fixed. For this, we used $L=$16, 20, 24 and 28 lattices.  At any finite
$L$, the additive renormalization of fermion mass is taken care of by
tuning $M_P$. 
Since
QED$_3$ is super-renormalizable, $M_P L$ is a finite additive renormalization.
We tabulate
the simulation parameters for the $N_f=1$ data set in \tbn{nf1data}.
We also show $M_P L$ as a function of $\ell$ graphically in \fgn{mw}. 
We used the same set of $\ell$ for $N_f=$ 2, 3 and 4 as well.

In the HMC simulation, we kept the molecular dynamics step-size $\Delta
t$ to be $1/N_{\rm MD}$. We tuned the number of steps per trajectory,
$N_{\rm MD}$, at run time to keep the Monte Carlo acceptance near 80\%. At
all simulation points, we ran about 13,500 trajectories with the first
300 trajectories discarded for thermalization. Then, we used only the
gauge configurations separated by an autocorrelation time, $\tau$, as
determined from the smallest eigenvalue of the Dirac operator.

\bibliography{biblio}
\end{document}